\documentclass{JHEP3}
\usepackage{amsmath}
\usepackage{graphicx}

\title{Swiss Cheese and a Cheesy CMB}
\author{Wessel Valkenburg\\
LAPTH\footnote{Laboratoire d'Annecy-le-Vieux de Physique
Th\'eorique, UMR5108}, Universit\'e de Savoie and
CNRS\\
BP110, F-74941 Annecy-le-Vieux Cedex\\
France\\
E-mail: \email{wessel.valkenburg@lapp.in2p3.fr}
}

\preprint{LAPTH-1318/09}

\abstract{It has been argued that the Swiss-Cheese cosmology can mimic Dark Energy, when it comes to the observed luminosity distance-redshift relation.  
Besides the fact that this effect tends to disappear on average over random directions,
we show in this work that based on the Rees-Sciama effect on the cosmic microwave background (CMB), the Swiss-Cheese model can be ruled out if all
holes have a radius larger than about 35 Mpc. We also show that for smaller holes, the CMB is not observably affected, and that the small holes can still mimic Dark Energy, albeit in special directions, as opposed to previous conclusions in the literature.
However,
in this limit, the probability of looking in a special direction where
the luminosity of supernovae is sufficiently supressed becomes very small, at least
in the case of a lattice of spherical holes considered in this paper.}

\begin{document}

\maketitle

\section{Introduction}
Ever since it was observed that distant supernovae (SN), of type Ia, appear dimmer than expected in a matter dominated spacetime~\cite{Riess:1998cb,Astier:2005qq}, when combined with 
measurements of the local Hubble factor~\cite{Freedman:2000cf}, cosmologists are led to the conclusion that the recent expansion of the universe definitely accelerates. Observations of the Cosmic Microwave Background (CMB) prefer a large angular diameter distance to $z=z_{\rm dec}\sim1100$, which in $\Lambda$CDM indicates a close to spatially flat universe~\cite{Smoot:1992td,Kuo:2002ua,Mason:2002tm,Komatsu:2008hk,Rasanen:2008be}. Given constraints on the matter content of the universe, from CMB and Large Scale Structure, this is achieved when a Cosmological Constant is present, given the assumption that the locally observed Hubble factor is equal to the global Hubble factor. The amount of clustering of galaxies is slowed down by accelerated expansion with respect to a pure cold dark matter universe~\cite{Efstathiou:2001cw}. The latest addition to this impressive evidence are the observed Baryon Acoustic Oscillations (BAO), which provide an observation of one length scale at different redshifts, and thereby measure the expansion over different times~\cite{Eisenstein:2005su,Percival:2007yw}.

Presented as such, the evidence for acceleration is convincing, see Ref.~\cite{Seikel:2007pk} though for critique. However, a number of assumptions needs to be made in order to come to the conclusion that a Cosmological Constant or a form of Dark Energy (DE) is there. If the locally observed Hubble factor cannot be extrapolated to the global expansion rate, as is the case if, {\em e.g.}, the observer lives in a large void, the observed angular diameter distance of SN may be explained without DE~\cite{Celerier:1999hp,Tomita:2000rf,Tomita:2000jj,Tomita:2001gh,Iguchi:2001sq,Tomita:2002df,Moffat:2005yx,Moffat:2005zx,Moffat:2005ii,Alnes:2005rw,Mansouri:2005rf,Vanderveld:2006rb,Garfinkle:2006sb,Biswas:2006ub,Chung:2006xh,Alnes:2006uk,Caldwell:2007yu,Alexander:2007xx,GarciaBellido:2008nz,Clifton:2008hv,GarciaBellido:2008gd,Hunt:2008wp}. Spatial flatness would in that case be achieved with a matter density equal to the critical density, $\Omega_{\rm M} =1 $, fitting the CMB~\cite{Hunt:2008wp}. A globally small Hubble constant is needed in that case to explain the age of the universe~\cite{Krauss:1995yb}. The clustering of galaxies can also be caused by a hot dark matter component, such as neutrino's with a mass of $m_\nu\sim 0.5$ eV~\cite{Hunt:2008wp}. The BAO data are obtained as interpreted data, in the sense that one has to apply a fiducial cosmology to the data in order to abstract the BAO. To our knowledge this has only been done assuming an FLRW-universe, and it is unclear how the BAO would be affected by applying different fiducial models. 

There is an ongoing debate as to whether today's universe is properly described by a perturbed FLRW-metric, assuming large scale homogeneity and isotropy, or not. See Refs.~\cite{Biswas:2007gi, Kolb:2008bn} for explications why the usual argument, that the universe today is everywhere described by a small Newtonian potential, may not be sufficient to rule out the role of inhomogeneities. If the universe is properly described by the FLRW-metric, then the evidence for DE is compelling. If it is not, then we need to understand why we observe an apparent acceleration, which is phenomenologically well described by $\Lambda$CDM. Many works have been devoted to possibilities linking the apparent acceleration to structure formation and to a departure from homogeneity of the Universe, none of them however convincing the community that structure formation explains the observations without the need for DE.  See Refs.~\cite{Ellis:2005uz,Rasanen:2006kp,Buchert:2007ik,Kolb:2009rp,Kundt:2009ky} for reviews. 

Here we will focus on one such proposition: the Swiss-Cheese Cosmology~\cite{1969ApJ...155...89K,1976ApJ...208L...1W,Sugiura:1999fm,Kozaki:2002ka,Brouzakis:2006dj,Marra:2007pm,Biswas:2007gi,Marra:2007gc,Bolejko:2008xh,Vanderveld:2008vi,Brouzakis:2007zi,Brouzakis:2008uw,Gurzadyan:2008yx,Gurzadyan:2008hj,Ghassemi:2009ug,Clifton:2009nv}. The Swiss-Cheese Cosmology is described globally by an FLRW-metric, but locally contains holes (voids surrounded by a mass shell) described by a Lema\^itre-Tolman-Bondi metric (LTB). The holes match flawlessly to the FLRW-metric at the borders, and are mass compensated in such a way that, from outside such a hole, the global effect of a patch containing such a hole is as if it were FLRW (cheese), due to Birkhoff's theorem~\cite{Einstein:1945id}. The Swiss-Cheese%
\footnote{The author must express his doubts about the name `Swiss Cheese'. The name implies that only {\em swiss} cheese contains holes, and at the same time that {\em all} swiss cheese contains holes.}  
toy model is an exact solution to the Einstein equations, hence does not suffer from averaging problems. Its goal is to approximate today's inhomogeneous universe in an exactly solvable manner, with most of the matter in structures, separated by voids. Recently it was shown that if curvature is a function of space, large local curvature at low redshift,  as in this model, is hardly constrained by observations~\cite{Clifton:2009kx}.  

The general idea of the Swiss-Cheese Cosmology is that photons travel through holes and structures, where holes have a lensing effect such that distant supernovae, observed through a number of holes, appear dimmer than in a homogeneous EdS Universe. This effect has been explored in the literature
and it was found:
\begin{itemize}
\item that if holes are perfectly aligned, a lensing effect can mimic a DE of about $\Omega_{\rm DE} = 0.4$, 
\item however, that averaged over many random distributions of holes, the effect vanishes, given that the holes are spherical.
\end{itemize}

In this work, we will show that, besides the on average disappearing of the lensing effect, the Swiss-Cheese Cosmology leaves a significant imprint on the CMB. It has been claimed that the effect that the holes have on the redshift of photons is marginal in comparison with the redshift at which the distant supernovae are observed. This reasoning does hold, but the effect is significant when compared to the observed anisotropies in the CMB-temperature, which are of the order $\mathcal{O}(10^{-5})$. For the first time we will show full sky CMB-maps in a Swiss-Cheese Cosmology, 
for different realisations, each realization with a different constant size of holes, $r_{\rm hl}$.
When showing these maps, we neglect the primordial power spectrum of the CMB-anisotropies, as the secondary anisotropies in most of the cases are overwhelmingly larger. These anisotropies are due to the Rees-Sciama~\cite{Rees:1968zz,Granett:2008xb,Masina:2008zv} effect, a non-linear late integrated Sachs-Wolfe effect due to structure formation.

At the same time, we will show full sky maps of the angular diameter distance, $d_{\rm A}$, at a fixed redshift in each direction. We confirm the results of Ref.~\cite{Vanderveld:2008vi}, that averaged over all directions, the angular diameter distance will appear as if the universe were exactly EdS, at least when the holes are spherical.
This leads to the conclusion that a Swiss-Cheese model that has predominantly holes with a radius larger than $35$ Mpc is ruled out, since it either cannot explain the supernovae in {\em every} direction, or it will leave a significant Rees-Sciama imprint on the CMB that is already ruled out by observations.

Our conclusion only applies to cosmologies in which the universe contains a very large number of large voids. In principle a smaller number of large voids is not ruled out by our analysis. In such a case, however, the voids would play no role at all in explaining the distant supernovae, which is not part of the scope of this paper.

We find that, in the density profile we consider, the maximal effect that the structure can have of the luminosity-distance-redshift relation is only marginally dependent on the size of the holes. This is a different conclusion than the one drawn in Refs.~\cite{Biswas:2007gi,Brouzakis:2008uw}. We will briefly address this difference.

The different dependence on the size of holes, for different choices of density profile, leaves the door open to an `Apollonian Gasket-like configuration'\footnote{The Apollonian Gasket is a fractal, constructed by starting out with three tangent circles, and subsequently filling areas between those circles with circles tangent to the previously drawn circles. See Ref.~\cite{BMandelbrot:1983}, and figure 1 in~\cite{Marra:2007pm}. }, in which the universe on average is FLRW, but locally is FLRW nowhere. In that case, an observer would see through holes in all directions. The CMB, if the configuration is such that all holes are sufficiently small, would be left in agreement with established perturbation theory. The holes can not be spherical, though, as the cancellation on average of the lensing effect~\cite{Vanderveld:2008vi} still applies to tiny holes as well. Probing such a configuration goes beyond the scope of this work,
as the model would need significant changes: the holes must be typically smaller than $r_{\rm hl}\sim 35$ Mpc, they must not be spherical, and they must be such that in all directions the chance of looking through holes is higher than looking through cheese. Altogether this poses serious difficulties for the Swiss-Cheese model. Besides, one might question if such a configuration approaches nature in any way, and if the effect remains significant with more natural density profiles.

In section~\ref{sec:model} we will briefly overview the metric and related equations describing the Swiss-Cheese cosmology and we will present the techniques used to calculate the full-sky maps of temperature and angular diameter distance. In section~\ref{sec:cmb} we present the full-sky CMB in a Swiss-Cheese universe, for different typical hole sizes. In section~\ref{sec:da} we will discuss the overall effect of the Swiss Cheese on distance measures.
We conclude in section~\ref{sec:conclusion}.

\section{The Model}\label{sec:model}
\subsection{The metric and geodesics}
\paragraph{The metric}
We define the Swiss-Cheese model identical to the model in Ref.~\cite{Marra:2007pm}.
In the Swiss-Cheese model, the metric is anywhere of the LTB form
\begin{align}
ds^2&=-dt^2+\frac{Y'^2(r,t)}{W^2(r)}dr^2+Y^2(r,t)d\Omega^2,
\end{align}
where $Y'(r,t)\equiv\partial_r Y(r,t)$. 
We can choose the coordinate system such that the universe is divided in equally sized cubic boxes, with sides of length $2r_{\rm ch}$, each with the origin of its own coordinate system in the center of that box. The time coordinate $t$ is scaled such that $t_0=0$ corresponds to today, and $t=-1$ corresponds to the time of the Big Bang.

In the cheese, we recover the FLRW-metric by setting $Y_{\rm ch}(r,t)=r a_{\rm ch}(t)$ and $W^2_{\rm ch}(r)=1-kr^2$. In the holes, the metric is determined by 
\begin{align}
\dot Y_{\rm hl}(r,t)=\sqrt{\frac{M(r)}{3 \pi Y_{\rm hl}(r,t)}+2E(r)},
\end{align}
with initial conditions at time $\bar t=-0.8$, 
\begin{align}
 W^2_{\rm hl}(r)-1\equiv E(r) =& \frac{1}{2}H^2_{\rm FLRW}(\bar t) r^2 - \frac{1}{6 \pi}\frac{M(r)}{r},\label{eq:wr}\\
 M(r)\equiv &4\pi \int_0^r \rho(u) Y_{\rm hl}^2(u,\bar t)Y_{\rm hl}'(u, \bar t) du,\\
 \rho(r)=&\left\{\begin{tabular}{ll}$A e^{-\frac{(r-r_{\rm M})^2}{2 \sigma^2}}+\epsilon $ & for $r<r_{\rm hl}$,\\&\\$\rho_{\rm ch}$& for $r>r_{\rm hl}$,\end{tabular}\right. \\
 Y(r,\bar t)=&r.
 \end{align}
 
In this setup, the free parameters are those describing the matter distribution in the holes: the size of the spherical LTB-metric $r_{\rm hl}$, the minimum density $\epsilon$, the energy density scale $A$, the comoving radius at which the peak of the mass distribution resides $r_{\rm M}$, the comoving width of the mass shell $\sigma$, and the FLRW-energy density $\rho_{\rm ch}$. The mass distribution is constant in time, and a function of comoving radius $r$ only. In physical coordinates, however, mass will be moving outwards to form a shell close to the border of the LTB-patch.

The LTB-evolution is `switched on' manually at time $\bar t=-0.8$, such that at $-1<t<-0.8$ spacetime is described 
by the homogeneous and isotropic FLRW-metric everywhere. In this way, the energy density and the scale factor are described by continuous functions of time everywhere. The spatial curvature $W(r)$, Eq.~\eqref{eq:wr}, however, appears instantaneously at time $t=\bar t$. This discontinuity plays no role in the quantities investigated here, henceforth it is taken for granted in this toy model.

For the derivation of and motivation for the choices made above, we refer the reader to Ref.~\cite{Marra:2007pm}.

\paragraph{Geodesic equations}
As a consequence of the spherical symmetry of each coordinate patch, any geodesic will lay in a spatial plane. Hence, without loss of generality, we can write the geodesic equations as four independent equations, as in Refs.~\cite{Brouzakis:2006dj,Marra:2007pm},
\begin{align}
\frac{dz}{d\lambda}=&\,-\frac{\dot Y'(r,t)}{Y'(r,t)}\left((z+1)^2-\frac{c_\phi^2}{Y^2(r,t)}\right)-c_{\phi}^2\frac{\dot Y(r,t)}{Y^3(r,t)},  &z(0)=&\,0, \label{eq:geo1}\\
\frac{dt}{d\lambda}=&\,z+1,  &t(0)=&\,0,\\
\frac{dr}{d\lambda}=&\,\frac{W(r)}{Y'(r,t)}\sqrt{(z+1)^2-\frac{c_{\phi}^2}{Y^2(r,t)}},  &r(0)=&\,r_{\rm obs},\\
\frac{d\phi}{d\lambda}=&\,\frac{c_\phi}{Y^2(r,t)} , &\phi(0)=&\,\phi_{\rm obs}\label{eq:geo2},
\end{align}
where $r_{\rm obs}$ and $\phi_{\rm obs}$ define the location of the observer. 

The constant $c_\phi$ is defined by the physical angle between the photon geodesic and a radial geodesic pointing (which itself has $c_\phi=0$) towards the photon geodesic. At any time in any point, the constant $c_\phi$ satisfies the relation,
\begin{align}
\cos \alpha &= g^i(y) x^i (y) g_{ij}(y),\\
c_\phi &= (1+z)Y(r,t)\sin\alpha ,
\end{align}  
where $\vec g$ denotes the spatial direction of the geodesic at point $y$, and $\vec x$ denotes the spatial part of a radial geodesic pointing to coordinate $y$.

\subsection{Dimensions and configurations}
\begin{figure}
\includegraphics[width=\textwidth]{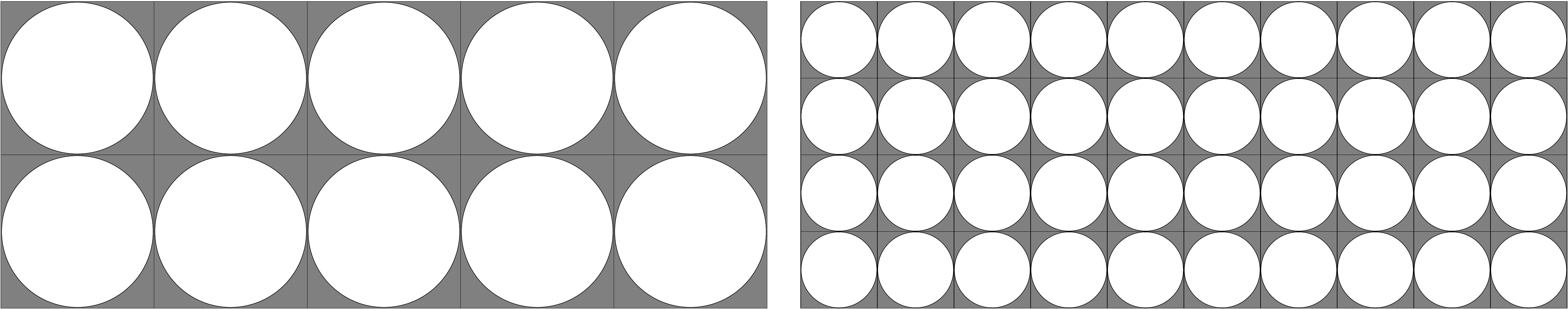}
\caption{A two-dimensional illustration of the configuration of holes throughout the cheese. Gray corresponds to the FLRW-metric, white corresponds to the LTB-metric. The difference between the left and the right illustration is $\kappa_{\rm left} = 2 \kappa_{\rm right}$. Throughout this work we will always have one size for all holes within each realization of a Swiss-Cheese universe.}\label{fig:allholes}
\end{figure}

We chose the parameters as in Ref.~\cite{Marra:2007pm}, being $r_{\rm ch}=r_{\rm hl}= 0.042 \kappa$, $\epsilon=0.0025$, $\sigma=r_{\rm hl}/10$, $r_{\rm M}= 0.037 \kappa$, $A=50.59$ and $\rho_{\rm FLRW}=25$, with however a freedom to chose the rescaling factor $\kappa$. For any $\kappa$, in physical dimensions these numbers correspond to holes with a radius of $350\kappa$ Mpc, in cubes with sides of length $l = 2 r_{\rm ch} = 700 \kappa$ Mpc, and $5/\kappa$ holes between the observer and $t=-0.8$ in an optimal direction, as illustrated in figure~\ref{fig:allholes}. The cheese is chosen to be spatially flat, with $\Omega_{\rm M}$=1, {\em i.e.}, EdS. We use units in which $c=16\pi G_{\rm N}=1$. The translation from these dimensions to physical dimensions is given in Table~\ref{tab:units}. Given the freedom to rescale, the choice of normalization is arbitrary. Throughout this work we will always have one size for all holes on a regular lattice within each realization of a Swiss-Cheese universe. Secondly, we will always place the observer in the cheese, at a spot where two adjacent holes are closest to each other. There is no particular reason to choose this location, other than for simplicity, and this choice is not important for the conclusions.

\begin{table}
\begin{tabular*}{\textwidth}{@{\extracolsep{\fill}}lllr}
Quantity          & Notation    & Unit            & Value            \\ 
\hline
mass density & $\rho(r,t)$, $\bar{\rho}(r,t)$ & $\rho_{C 0}$ 
& $9.2\times10^{-30}\textrm{ g cm}^{-3}$             \\
time              & $t$, $T$, $\bar{t}$, $t_{BB}$, $T_0$ 
& $(6 \pi \rho_{C 0})^{-1/2}$  & $9.3\textrm{ Gyr}$ \\
comoving radial coordinate & $r$         & $(6 \pi \rho_{C 0})^{-1/2}$  
& $2857 \textrm{ Mpc}$ \\
metric quantity   & $Y(r,t)$    & $(6 \pi \rho_{C 0})^{-1/2}$  
& $2857 \textrm{ Mpc}$ \\
expansion rate    & $H(r,t)$    & $(6 \pi \rho_{C 0})^{1/2} $ 
& $\frac{3}{2}H_{0,\, Obs}$ \\
spatial curvature term    & $W(r)$      & $1$     &        ---             \\
\end{tabular*}
\caption{\label{tab:units} The units used throughout this work, where $c=16\pi G_{\rm N}=1$. The present critical density is
$\rho_{C0}=3H^{2}_{0,\, Obs}/8 \pi$, with $H_{0,\, Obs}=70 \textrm{ km
s}^{-1}\textrm{ Mpc}^{-1}$. Table taken from Ref.~\cite{Marra:2007pm}}
\end{table}

\subsection{Different methods for the angular diameter distance}

The angular diameter distance, $d_{\rm A}$, is defined as the ratio of the size of an object at some distance of an observer, and the angular diameter at which it is observed. When, in the configuration of aligned bubbles, an observer is located in the cheese at a point where two adjacent bubbles are closest to each other, and the observer sees a source for which the geodesic connecting the observer and the center of the source is a straight line exactly through the centers of a number of bubbles, the angular diameter distance to that source amounts to
\begin{align}
	\tilde d_{\rm A}(\lambda) &= \frac{ Y(r(\lambda),t(\lambda)) \sin \phi(\lambda)}{\alpha}, \label{eq:simpleda}
\end{align} 
with $\alpha$ the small angle between the radial geodesic and a geodesic from an edge of the source to the observer, the latter geodesic described by $r(\lambda)$, $t(\lambda)$ and $\phi(\lambda)$~\cite{Marra:2007pm}. For small $\alpha$ this quantity is independent of $\alpha$. In an actual (numerical) calculation, one can integrate the geodesic equations for successively smaller alpha, until the quantity $\tilde d_{\rm A}$ converges to the same number along the geodesic. 

Eq.~\eqref{eq:simpleda} obviously only holds for the special case of the aforementioned purely radial geodesic connecting source and observer. The same method could be used in any direction, shooting two photons in almost but just not the same direction. This method would be computationally expensive though, since for each direction one has to reintegrate several times the geodesic equation, at very high accuracy such that the relative error in the tiny displacement orthogonal to the direction of the photon is small. The orthogonal displacement itself will already be extremely small compared to the distance travelled along the geodesic, given the small angle at which one shoots. 

In stead, equations describing the exact beam size along the geodesic are, as in \cite{Brouzakis:2006dj},
\begin{align}
	\frac{d\theta}{d\lambda}&=-\frac{2}{3} \rho(r,t) \left(\frac{dt}{d\lambda}\right)^2 - \theta^2 - \sigma^2\\
	\frac{d \sigma}{d \lambda} + 2 \theta \sigma &=  \frac{2}{3} \left(\frac{d\phi}{d\lambda}\right)^2 Y(r,t)^2
 		\left(\rho(r,t) - \frac{3 M(r)}{4\pi Y(r,t)^3} \right)\\
	\frac{1}{\sqrt{A}}\frac{d^2\sqrt{A}}{d\lambda^2}&=-\frac{2}{3}\rho(r,t) \left(\frac{dt}{d\lambda}\right)^2 - \sigma^2.
\end{align}
Here $A$ denotes the beam size, $\theta$ is the beam expansion, defined through $\theta = \frac{1}{2A}\frac{dA}{d\lambda}$, and $\sigma$ denotes the beam shear. The beam stretching becomes $\theta \pm \sigma$ in two orthogonal directions. We refer the reader to Refs.~\cite{Brouzakis:2006dj,Sachs:1961zz} for further explanations. The angular diameter distance $d_{\rm A}$ and the luminosity distance $d_{\rm L}$ to an observer at any point along a geodesic pointing from / to the observer then become,
\begin{align}
	d_{\rm A}(\lambda) &= \sqrt{\frac{A(\lambda)}{\Omega_{\rm source}}}\\
	d_{\rm L}(\lambda) & = (1+z)^2 d_{\rm A}(\lambda),
\end{align}
with $\Omega_{\rm source}$ the solid angle at which the source is observed. The initial conditions for a beam are,
\begin{align}
	\left.\frac{d\sqrt{A}}{d\lambda}\right|_{\lambda = 0}&=\sqrt{\Omega_{\rm source}},\\
	\left.\sqrt{A}\right|_{\lambda = 0}&= 0.\label{eq:inisqa}
\end{align}
With these initial conditions, $\theta$ and $\sigma$ are ill-defined at the initial conditions, as $\lim_{\lambda \rightarrow 0} \theta  = \infty.$ If we write $\xi = A \sigma$, we find the only relevant equations,
\begin{align}
	\frac{d\xi}{d\lambda}&=\frac{2}{3} A \left(\frac{d\phi}{d\lambda}\right)^2 Y(r,t)^2
 		\left(\rho(r,t) - \frac{3 M(r)}{4\pi Y(r,t)^3} \right)\label{eq:exactda1}\\
	\frac{d^2\sqrt{A}}{d\lambda^2}&=-\frac{2}{3}\sqrt{A}\,\,\rho(r,t) \left(\frac{dt}{d\lambda}\right)^2 - \frac{\xi^2}{A^{3/2}},\label{eq:exactda2}
\end{align}
with the extra initial condition
	$\left.\xi\right|_{\lambda = 0}= 0$,
if we demand that $\sigma$ is well behaved, {\em i.e.} finite, at all times. With these equations, the calculation of the angular diameter distance, and thereby the luminosity distance, amounts to simply including two more equations in the (numerical) integration scheme. Since $d_{\rm L} = (1+z)^2 d_{\rm A}$, hereafter we will mainly focus on $d_{\rm A}$ for simplicity.

We checked that both the intuitive, Eq.~\eqref{eq:simpleda}, and the exact, Eqs.~(\ref{eq:exactda1},~\ref{eq:exactda2}), method agree in the case where both can be applied.

\subsection{Using 2D geodesic equations in a 3D setup}

The spatial plane in which the photon's geodesic lays, is defined by the spatial part of a radial geodesic and the spatial part of the photon's geodesic. In the practice of this calculation, a photon exits one coordinate patch (A) and enters another patch (B) when one of its cartesian coordinated reaches $r_{\rm ch}$.
Let $\vec g$ denote the spatial direction of the geodesic in cartesian coordinates, let $\vec x^{\rm A}_0$ denote the cartesian coordinates at which the geodesic {\em entered} patch A, and 
let $\hat v\equiv\vec v / |\vec v|$ for any spatial vector $\vec v$. In cartesian coordinates, the direction $\vec g$ is invariant under translations, hence invariant under a transformation from one coordinate patch to another. This simplification only holds because the transformation is done at a point where spacetime is described by the spatially flat FLRW-metric. 
During a time step of integration of the geodesic equation, corresponding to photons leaving patch A, whilst moving in a plane spanned by the orthonormal vectors $\hat e^{\rm A}_1$ and $\hat e^{\rm A}_2$, the transformation at the border at time $\lambda_{\rm cross}$ is done as follows:
\begin{align}
\vec g^{\rm B}(\lambda_{\rm cross})&=\vec g^{\rm A}(\lambda_{\rm cross}),\\
\vec x^{\rm B}_0&= \vec x^{\rm A}_0+\vec e^{\rm A}_1 r_{\rm A}\cos \phi_{\rm A} + \vec e^{\rm A}_2 r_{\rm A}\sin \phi_{\rm A}\pm\left(\begin{tabular}{c} 0\\0\\$2 r_{\rm ch}$\end{tabular}\right),\\
\hat e_1^{\rm B}&=\hat x^{\rm B}_0,\\
\hat e_2^{\rm B}&=\frac{(\hat e_1^{\rm B} \cdot \hat g^{\rm B}) \hat e_1^{\rm B} - \hat g^{\rm B}}{|(\hat e_1^{\rm B} \cdot \hat g^{\rm B}) \hat e_1^{\rm B} - \hat g^{\rm B}|} ,\\
r^{\rm B} &= \left|\vec x^{\rm B}_0\right|,\\
\phi^{\rm B}&=0,\\
\cos \alpha^{\rm B} &= g^ix_{\rm B}^ig^{\rm B}_{ij},\\
c^{\rm B}_\phi &= (1+z)Y(r,t)\sin\alpha^{\rm B} .
\end{align}
where all quantities are evualated at $\lambda_{\rm cross}$. The integration then continues in box B with initial values  $r_{\rm B}$ and $\phi_{\rm B}$, defined in the plane spanned by the orthonormal vectors  $\hat e^{\rm B}_1$ and $\hat e^{\rm B}_2$. In this example we assumed that it is the third cartesian coordinate of vector $\vec g$ that hits the border of the patch. Figure~\ref{fig:match_schematic} gives a visual explanation of the relation of different vectors.

\newlength{\totfigwidth}
\setlength{\totfigwidth}{0.8\textwidth}
\begin{figure}
\begin{center}
\setlength{\unitlength}{0.001\totfigwidth}
\begin{picture}(0,0)(0,0)%
\includegraphics[width=\totfigwidth]{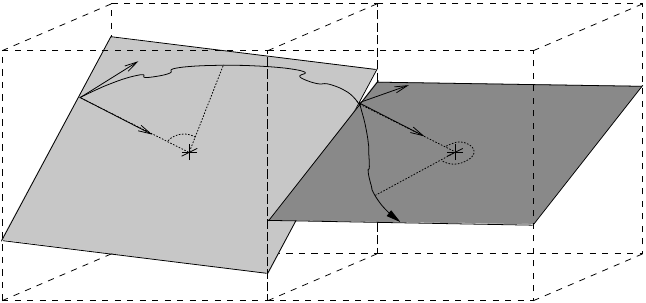}%
\end{picture}%
%
\begin{picture}(1000,466.15)(0,0)
\put(180,290){$\hat e_1^{\rm A}$}
\put(150,360){$\hat e_2^{\rm A}$}
\put(255,265){$\phi^{\rm A}$}
\put(320,295){$r^{\rm A}$}
\put(430,320){$\vec g^{\rm A}$}
\put(600,290){$\hat e_1^{\rm B}$}
\put(600,350){$\hat e_2^{\rm B}$}
\put(735,225){$\phi^{\rm B}$}
\put(620,170){$r^{\rm B}$}
\put(520,220){$\vec g^{\rm B}$}\end{picture}%
\end{center}
\caption{A cartoon of a geodesic in two adjacent patches. The geodesic is always in a plane in each patch, spanned by orthonormal vectors $\hat e_1$ and $\hat e_2$, both defined when entering a patch. At the exact border of the coordinate patches we have in three-dimensional cartesian coordinates $\vec g^{\rm B}(\lambda_{\rm cross})=\vec g^{\rm A}(\lambda_{\rm cross})$. }\label{fig:match_schematic}
\end{figure}
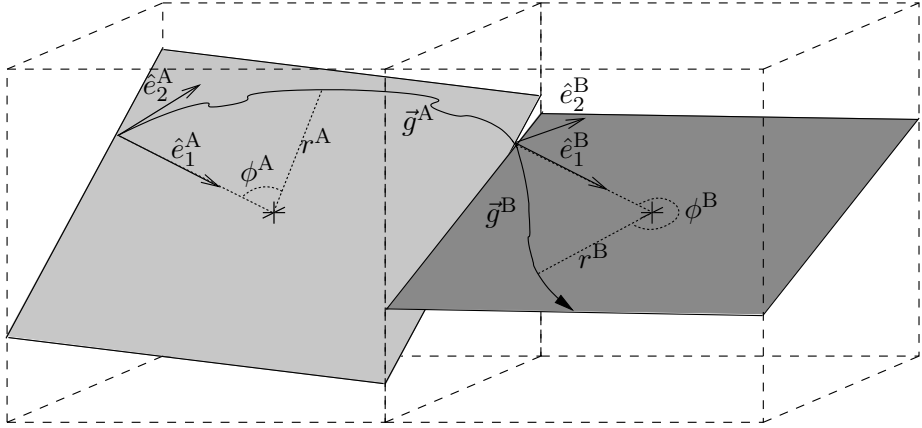

\section{The CMB}\label{sec:cmb}

\begin{figure}
\begin{center}
\begin{tabular}{lr}
\includegraphics[width=0.45\textwidth]{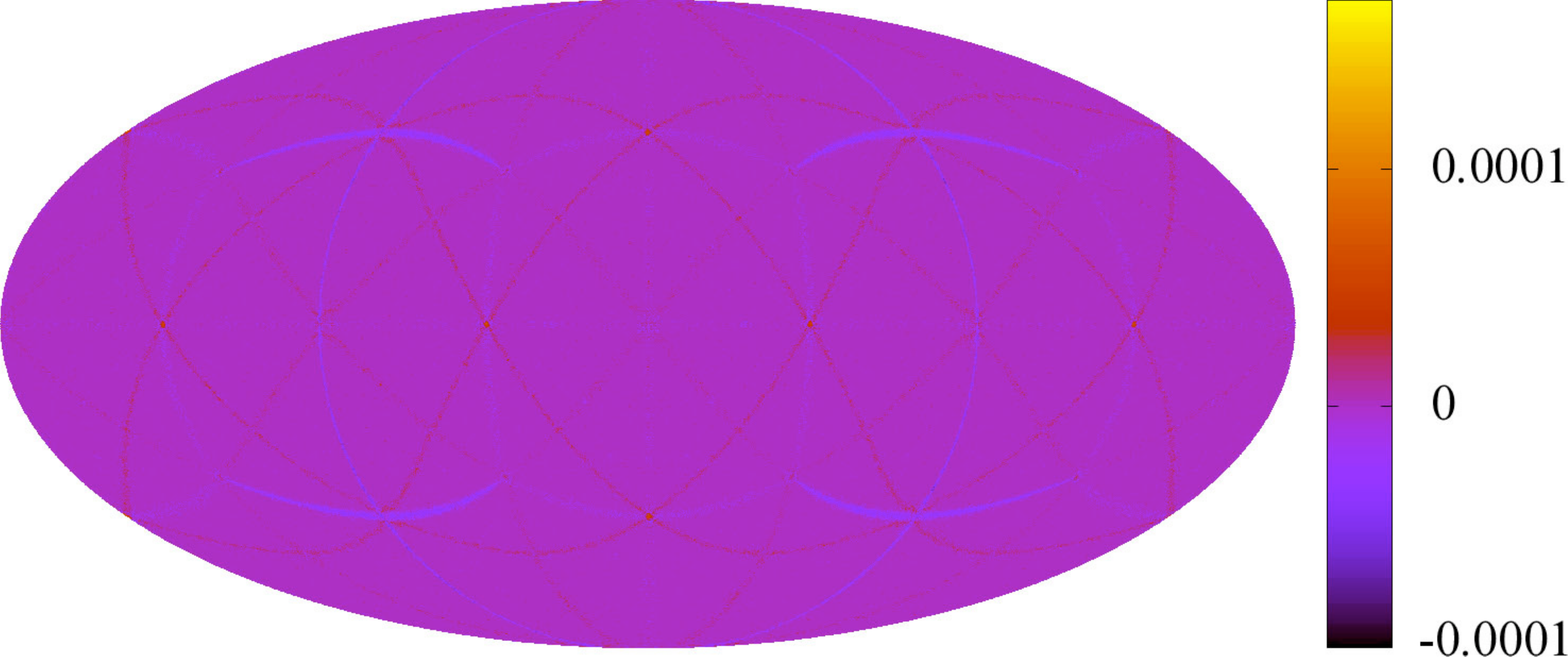}&
\includegraphics[width=0.45\textwidth]{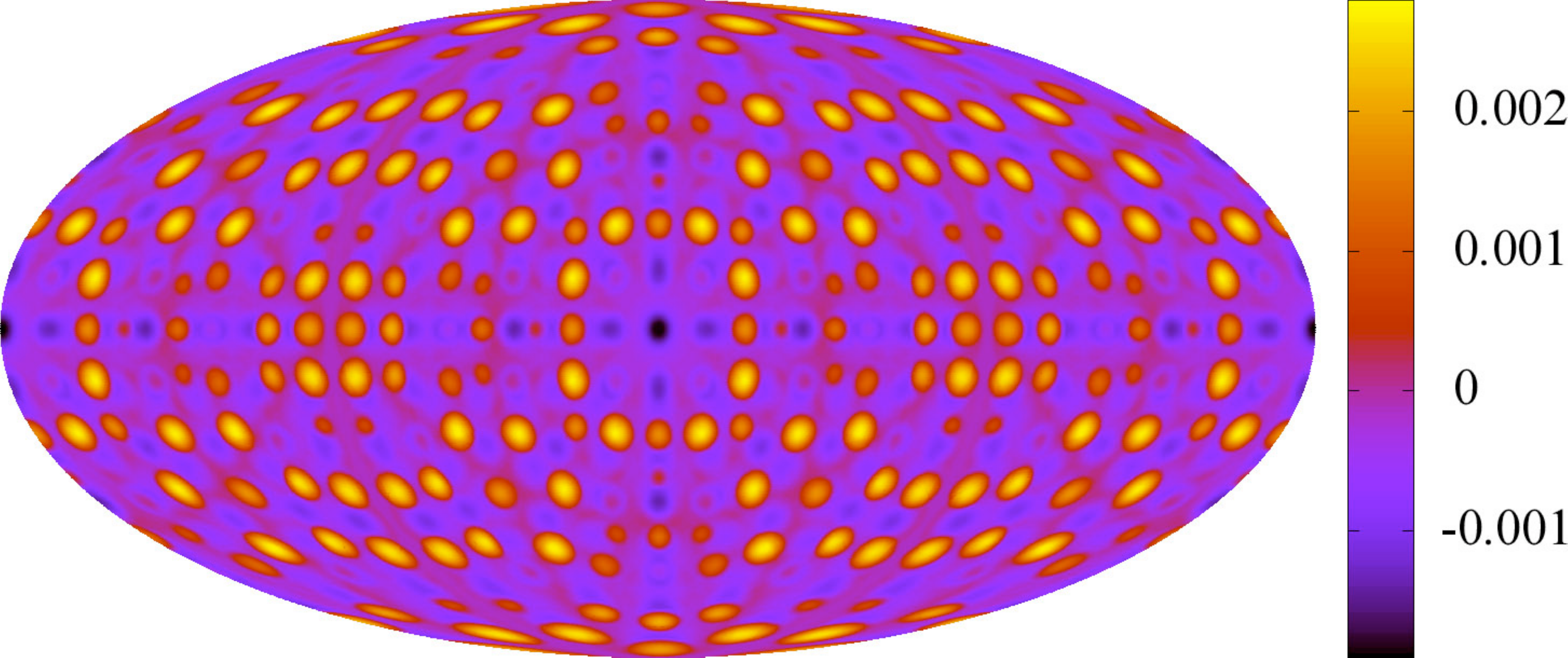}\\
\includegraphics[width=0.45\textwidth]{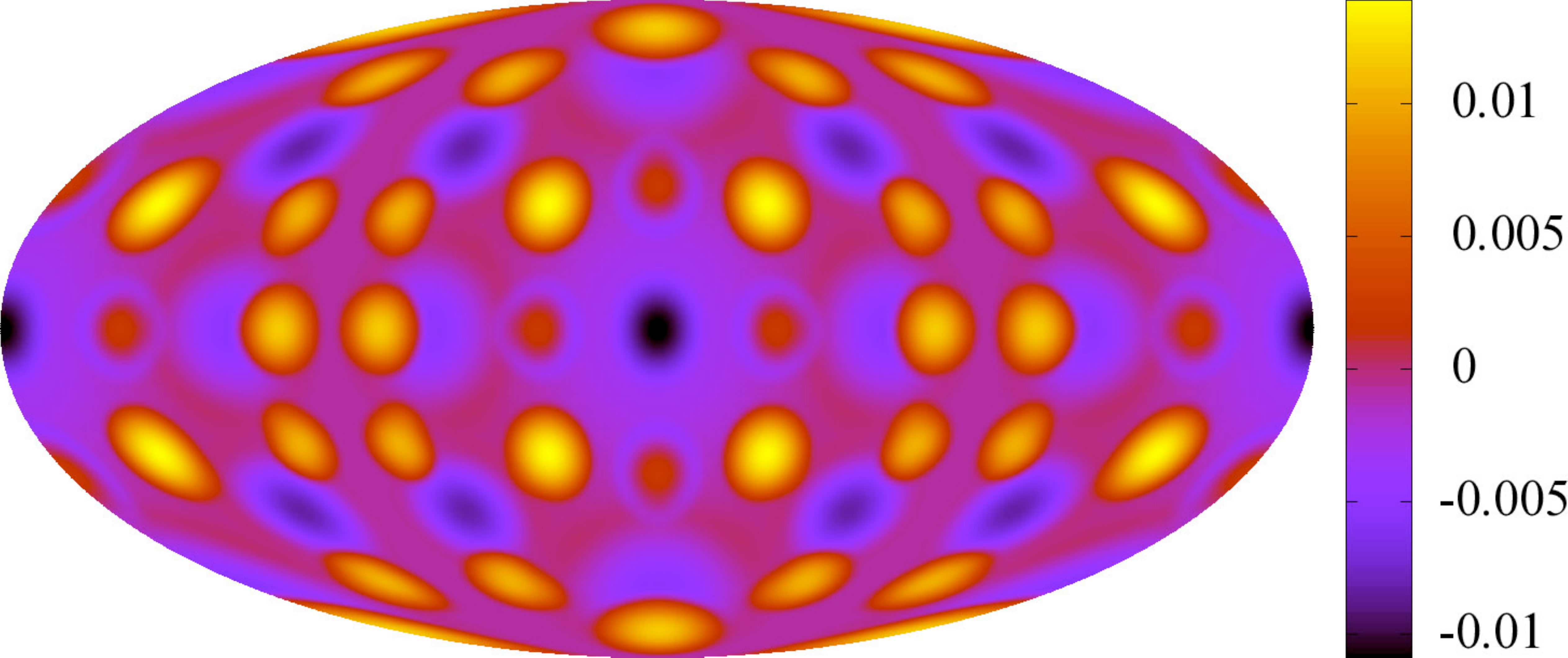}&
\includegraphics[width=0.45\textwidth]{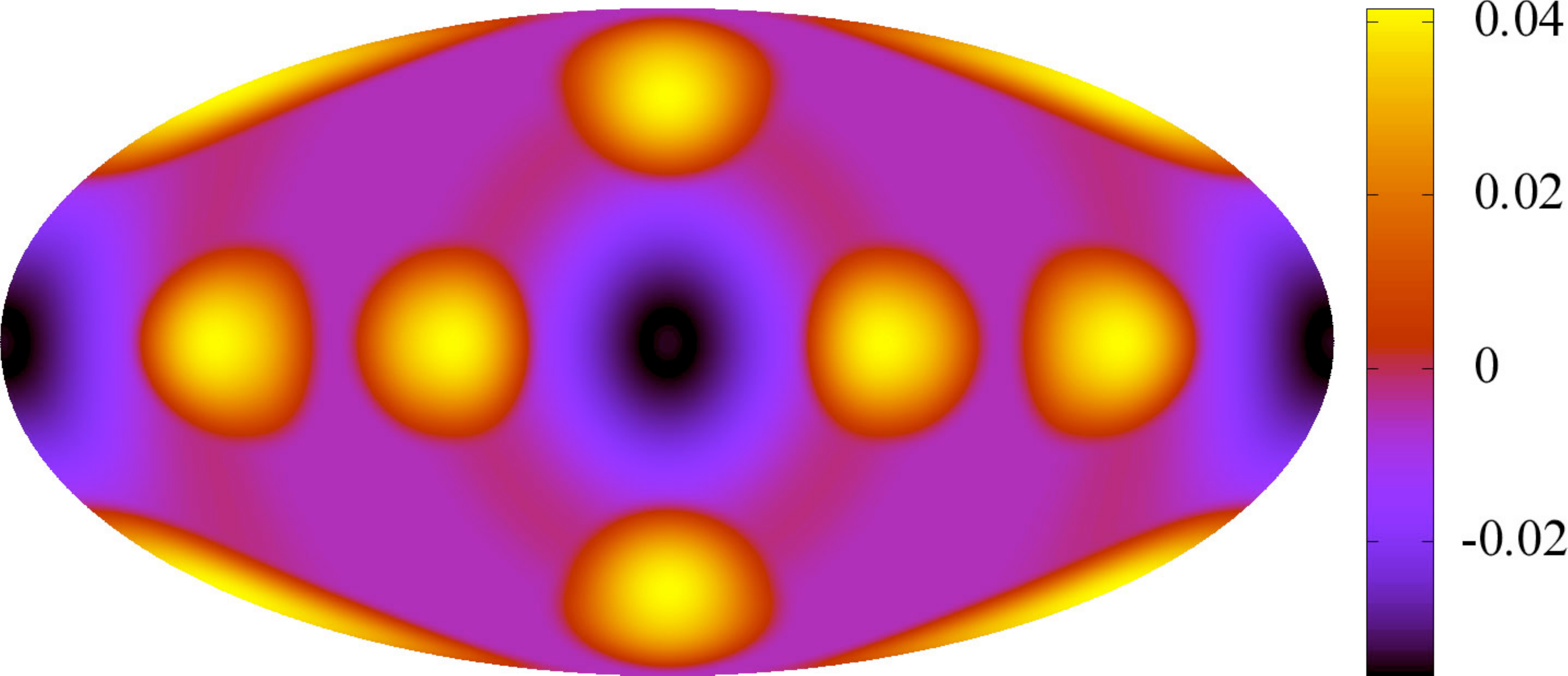}
\end{tabular}
\end{center}
\caption{Full-sky maps of secondary CMB-anisotropies induced by the Swiss-Cheese structure, in Mollweide-projection. The quantity shown is $\frac{T-\bar T}{\bar T}$. These temperature maps would be observed in Swiss-Cheese universes with holes of size $r_{\rm hl} =$ 3.5 Mpc (upper left figure), $r_{\rm hl} =$ 350 Mpc  (upper right figure), $r_{\rm hl} =$ 875 Mpc  (lower left figure) and $r_{\rm hl} =$ 1.75 Gpc  (lower right figure). The cold spots correspond to photons that experienced a large Rees-Sciama effect, hot spots correspond to photon's that traveled through a non-integer number of holes, and all ring-like structures are artifacts of the regular distribution of holes in this configuration.}\label{fig:cmb-cheese1}
\end{figure}

\subsection{Temperature maps and their power spectra} 
As the universe is described by the matter-dominated FLRW-metric at $-1<t<\bar t= -0.8 $, photons obtain no anisoptropies from the Swiss-Cheese structure in that period. It suffices to integrate photon geodesics backwards in time from today back to $t=\bar t$, and then to compare 
redshift
a photon has experienced with the 
redshift
experienced on average in all directions. This translates directly to a relative temperature difference, as $T \propto a^{-1} \propto 1+z$. Hence, shooting photons in all directions and obtaining the redshift at time $\bar t = -0.8$ as a function of direction, one obtains a temperature anisotropy map of the CMB with only secondary anisotropies, caused solely by the Swiss-Cheese matter distribution. 

We show full-sky CMB anisotropy maps for four different hole sizes in figure~\ref{fig:cmb-cheese1}.
The quantity shown is $\frac{T-\bar T}{\bar T}$, where $\bar T$ is an
average over all directions. 
In each direction, the photon departed at the same time $t=\bar t$, towards us.
The hole sizes for these maps are $r_{\rm hl} = 3.5$ Mpc (top left), $r_{\rm hl} = 350$ Mpc (top right),  $r_{\rm hl} = 875$ Mpc (bottom left) and  $r_{\rm hl} =  1.75$ Gpc (bottom right). 
\begin{figure}
\begin{center}
\includegraphics[angle=-90,width=.7\textwidth]{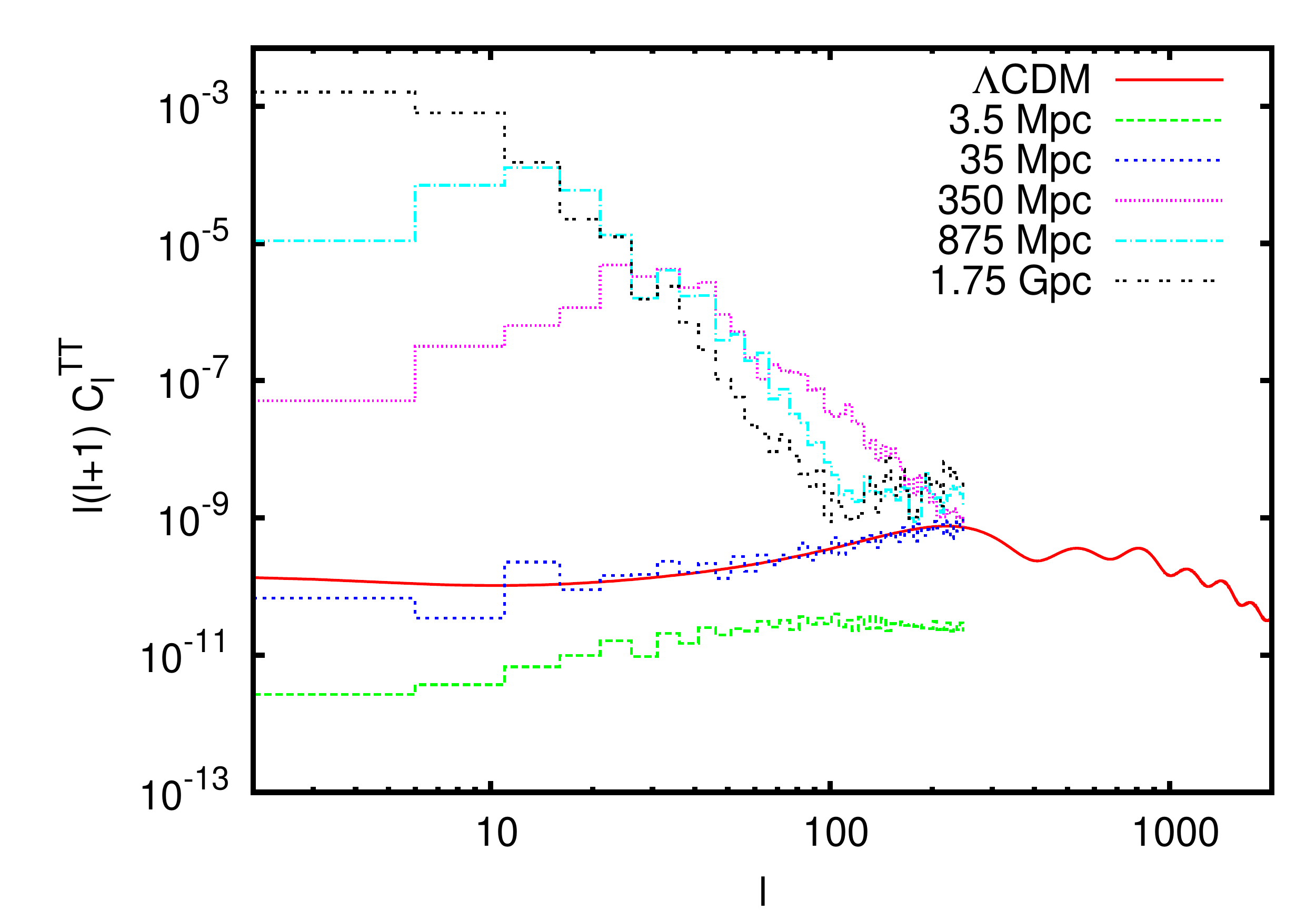}
\end{center}
\caption{The $C_l^{\rm TT}$ of secondary anisotropies for different Swiss-Cheese cosmologies, in bins of 5 multipoles. For comparison we plot the full $C_l^{\rm TT}$-spectrum in a standard $\Lambda$CDM cosmology with $\Omega_{\rm baryon} = 0.045$, $\Omega_{\rm cdm} = 0.245$, $\Omega_{\rm k} = 0$, $\Omega_{\rm DE} = 0.71$, $h=0.7$ (red solid line, unbinned). A Swiss-Cheese universe with holes of radius smaller than 35 Mpc, potentially leaves the CMB unaltered with respect to the standard cosmological model. Note the numerical limitations beyond the ankle at high $l \sim 100$, as discussed in the text.}\label{fig:cmb-cls1}
\end{figure}
The angular power spectra, $C_l$'s, of the CMB-temperature anisotropy autocorrelation in different configurations are displayed in figure~\ref{fig:cmb-cls1}. They have been calculated using the Healpix package~\cite{Gorski:2004by}. The power spectra displayed are defined in general for any quantity $X$ as,
\begin{align}
	a^X_{lm}&\equiv\int_0^{2\pi}d\phi\int_0^\pi d\theta\,\, \frac{X(\theta,\phi) - \bar X}{\bar X} \,\, Y_{lm} (\theta,\phi)\\
	C^{XX}_l&\equiv\frac{1}{2l+1}\sum_{m=-1}^{m=l}\left | a^X_{lm}\right |^2,
\end{align}
with $Y_{lm}$ being the spherical harmonics.

\subsection{Numerical limitations}
Before discussing the power spectra, a note about numerical limitations needs to be made.
We can only consider the power spectra up to the multipoles as displayed, because for even higher multipoles $l$, a much larger resolution in the maps would be needed. Even higher resolution would be computationally expensive, up to the point where numerical errors in the integration will dominate the anisotropy at such small scales. 

All spectra show an ankle at about the same multipole $l \sim 100$. The spectrum beyond this ankle cannot be trusted, as the ankle is already an aliasing artifact of the resolution of choice. When the resolution is increased, the ankle goes down and moves to higher $l$, for all five configurations displayed. The resolution chosen, $3 \times 10^6$ pixels per map, is at the limit of the computational resources at hand, with our code as is\footnote{The calculations were performed on an MPI-grid. For the largest holes, the computation of one map takes about one hour on 16 cpu's. For the tiniest holes, the computation of a map takes about 30 hours on 128 cpu's. }. 
 
\subsection{Discussion}\label{sec:cmb:disc}
By eye one already sees, in figure~\ref{fig:cmb-cheese1}, that the maximum temperature anisotropy is much larger than the observed average CMB-temperature anisotropy, notably for larger holes. The center of each map points in the special direction of exactly aligned holes. The cold spots are directions in which the photons travel through an optimal (read largest) number of holes. These are in fact cold, and not hot, spots because travelling through many holes, implies crossing many shells of in falling matter as well. What is observed here is actually the Rees-Sciama effect. The hot spots correspond to directions in which the photon at $t=\bar t$ was in a region that became a hole. {\em I.e.}, the photon has not travelled through an integer number of holes. 
When crossing an
entire hole, a photon passes both regions in which the matter, falling outwards to form a shell around the hole, falls against the direction of photon and regions in which the matter falls in the same direction as the photon. Any Rees-Sciama effect
mostly cancels in that case. When the photon is in the middle of a region that will become a hole when the matter starts falling, the photon automatically will only experience the metric of matter that falls in the direction of the photon path. In that hole, the cancellation will not happen, hence the photon relatively gains energy. 
 One might wonder wether these hotspots are natural, and if it would not make more sense to constrain the number of holes that a photon crosses to be integer, for $-0.8<t<0$. Firstly, if the Swiss Cheese is a toy model for structure formation, then there is nothing unnatural about the period when the matter perturbations undergo the transition from linear to non-linear perturbations. In the linear regime, the potentials do not change. But still, if the photon happens to cross such a potential well at the time of transition, the photon will become part of a similar hot spot as in this toy model. Secondly, the temperature deviation of the hot spots from the average temperature is about the same as the deviation of the cold spots. It is likely that removing the hot spots would only change the angular correlation by a factor of two. This should be investigated though, keeping in mind that by eye one cannot do much statistics on a picture. In this work we choose to stick to one well-defined model.

In each map one also observes the same circular ring of average temperature, centered on the map. In the projection and orientation chosen, this ring corresponds to the plane surrounding the observer in which the photons always travelled through cheese only. This is a direct consequence of the regular distribution of holes chosen. Other observed lines, especially visible in the map of smaller holes, are similarly artifacts of special directions in the regular distribution of holes.

Let us consider the angular power spectra, $C_l$'s, of the autocorrelation for these maps, in figure~\ref{fig:cmb-cls1}. The spectra are binned in bins of five multipoles. The Swiss-Cheese spectra must be considered as additional secondary anisotropies in the CMB, since the primary power spectrum has been ignored. For comparison the power spectrum of the CMB, as it would be observed today in the standard $\Lambda$CDM universe, is displayed. For holes of radius $r_{\rm hl} = 35$ Mpc, the secondary anisotropies are of the same magnitude as the CMB observed today. This means that a Swiss-Cheese universe with all holes smaller than 35 Mpc potentially leaves the CMB as observed unaltered. For example, the secondary anisotropies for holes of radius 3.5 Mpc are up to two orders of magnitude smaller than the observed anisotropies. A decrease of the Rees-Sciama effect with decreasing hole-size was also foreseen in Refs.~\cite{Marra:2007pm,Biswas:2007gi}. In a more realistic distribution of holes, the size of holes is not a fixed number. In that case, probably a few holes larger than $r_{\rm hl}\sim 35$ Mpc amongst smaller holes, would leave an imprint on the CMB small enough to agree with observations.

\section{Angular diameter distance - redshift relation}\label{sec:da}
\subsection{The same maximum distance for all hole sizes up to 1.75 Gpc}
\begin{figure}
\vspace{-0.5cm}
\begin{center}
\includegraphics[angle=-90,width=.47\textwidth]{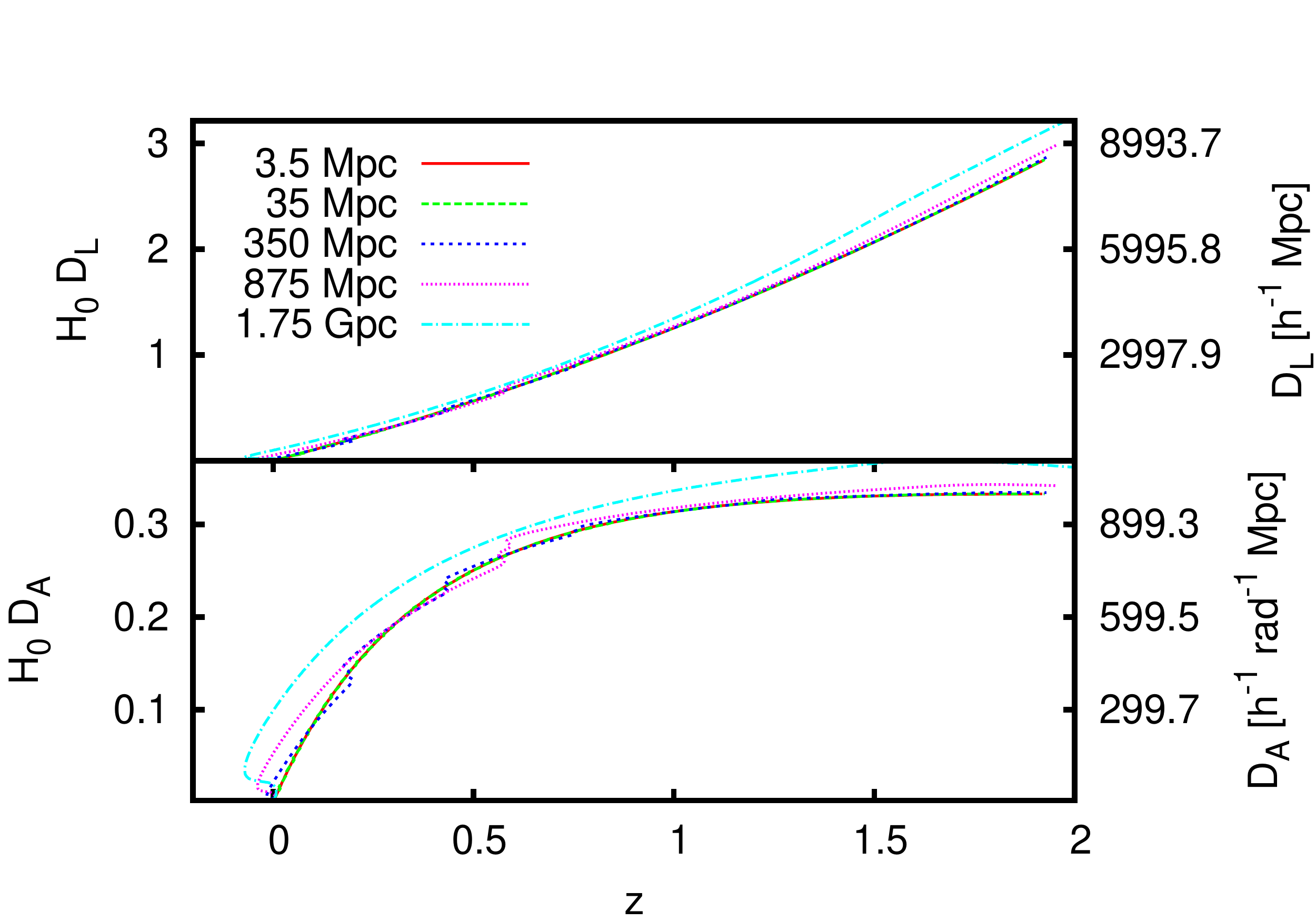}
\includegraphics[angle=-90,width=.47\textwidth]{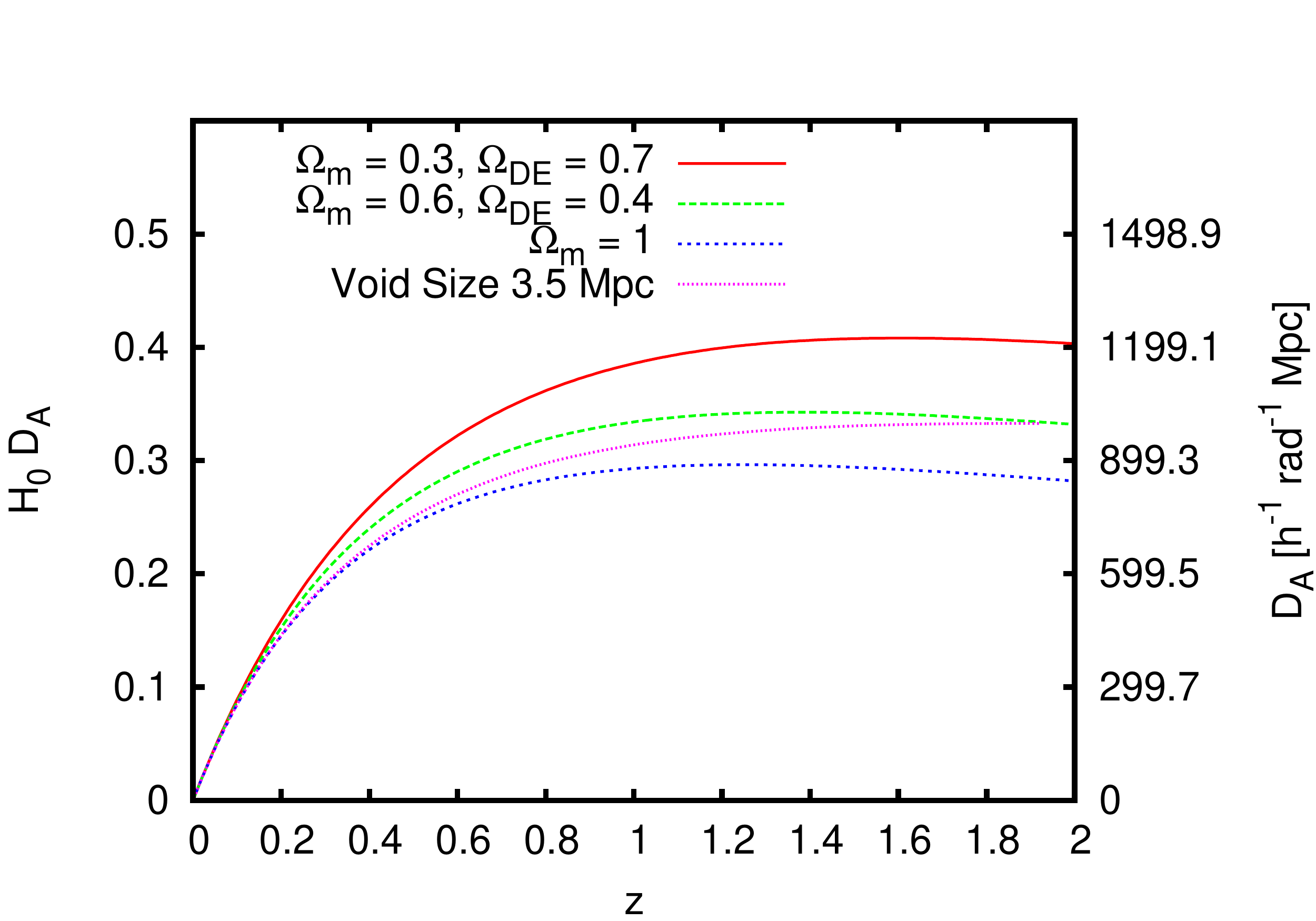}
\end{center}
\caption{{\em Left:} The maximum luminosity distance (top) and the maximum angular diameter distance (bottom) achievable as a function of redshift $z$, for Swiss-Cheese universes with five different hole sizes. These distances would be observed when looking through a series of perfectly aligned holes. Note that the maximum lensing effect only marginally changes with hole size. This indicates that the ratio between time spent in 'cheese' and time spent in 'holes' along the history of a photon path is the quantity that defines the observed distances, as opposed to the physical size of the holes. {\em Right:} The angular diameter distance as a function of redshift $z$, comparing the Swiss-Cheese model with three different spatially flat FLRW-solutions.}\label{fig:AnDL}
\end{figure}

In the right of figure~\ref{fig:AnDL}, we repeat for illustration a comparison of $d_A(z)$ for different FLRW-cosmologies and one particular Swiss-Cheese cosmology, similar to Ref.~\cite{Marra:2007pm}. Here we focus on the maximum deviation in $d_A$ with respect to an FLRW-cosmology with $\Omega_{\rm matter}=1$. At redshift $z\sim 1.92$, in this Swiss-Cheese cosmology  $d_{\rm A}$ corresponds to that in a flat FLRW-universe with $\Omega_{\Lambda}=0.4$. In Ref.~\cite{Marra:2007gc} it was argued, however, that a good fit over the whole curve to a cosmology with $\Omega_{\Lambda} = 0.75$, is actually achieved by a Swiss-Cheese universe with five holes of radius 250 Mpc between the observer and $t=\bar t$. Since we are dealing with a toy model that probes the ability of structure formation to explain the observed acceleration, we prefer to probe the maximum achievable effect at high redshift, over finding the best out of a set of perhaps mediocre fits.

In the left of figure~\ref{fig:AnDL}, the maximum $d_{\rm A}$ is compared for all models. The maximum $d_{\rm A}$ for all models lies in the direction of perfectly aligned holes. In nature such a direction is unlikely to exist, but a more natural distribution of holes goes beyond our scope here.
The maximum effect on $d_{\rm A}$ is only marginally dependent on the size of holes. 
This conclusion seems in contradiction with conclusions drawn in Refs.~\cite{Biswas:2007gi,Brouzakis:2008uw}, which we address at the end of this section.

\subsection{Distribution in different directions}
\begin{figure}
\begin{center}
\begin{tabular}{lr}
\includegraphics[width=0.45\textwidth]{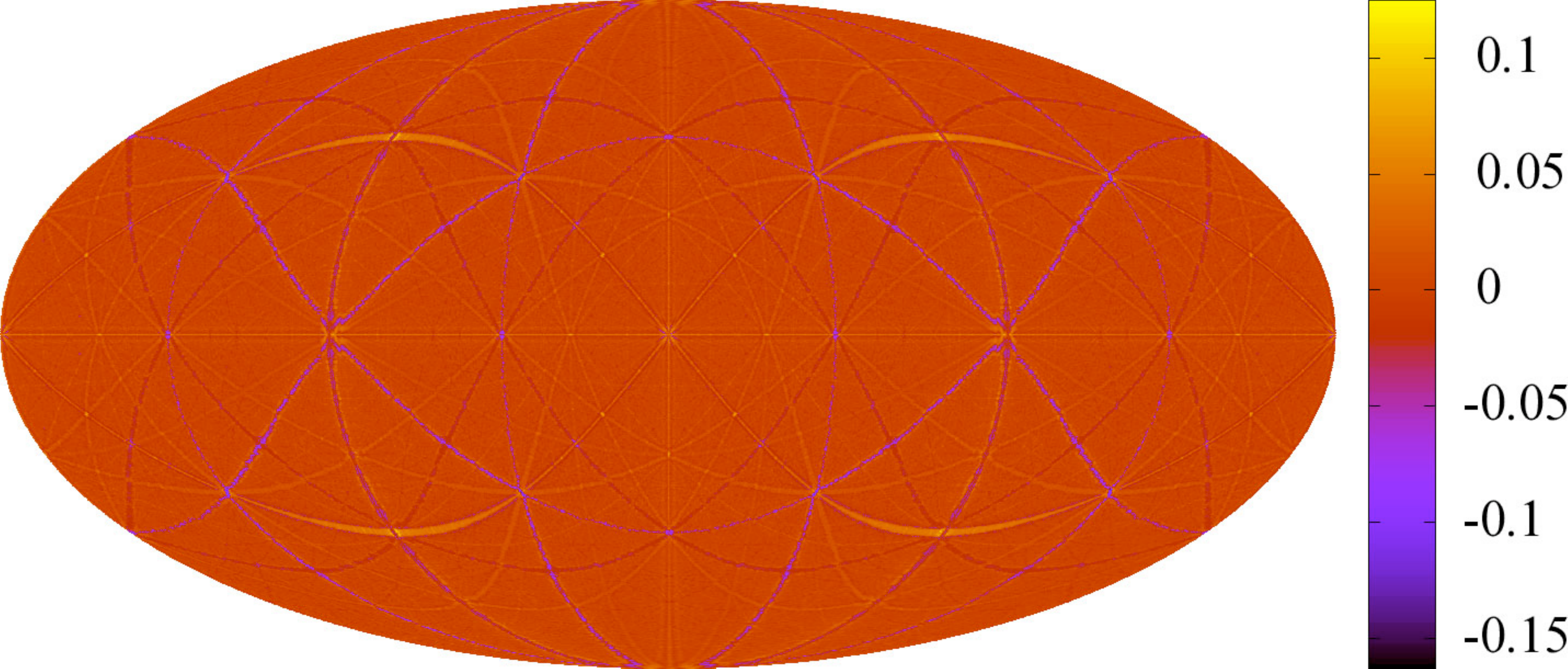}&
\includegraphics[width=0.45\textwidth]{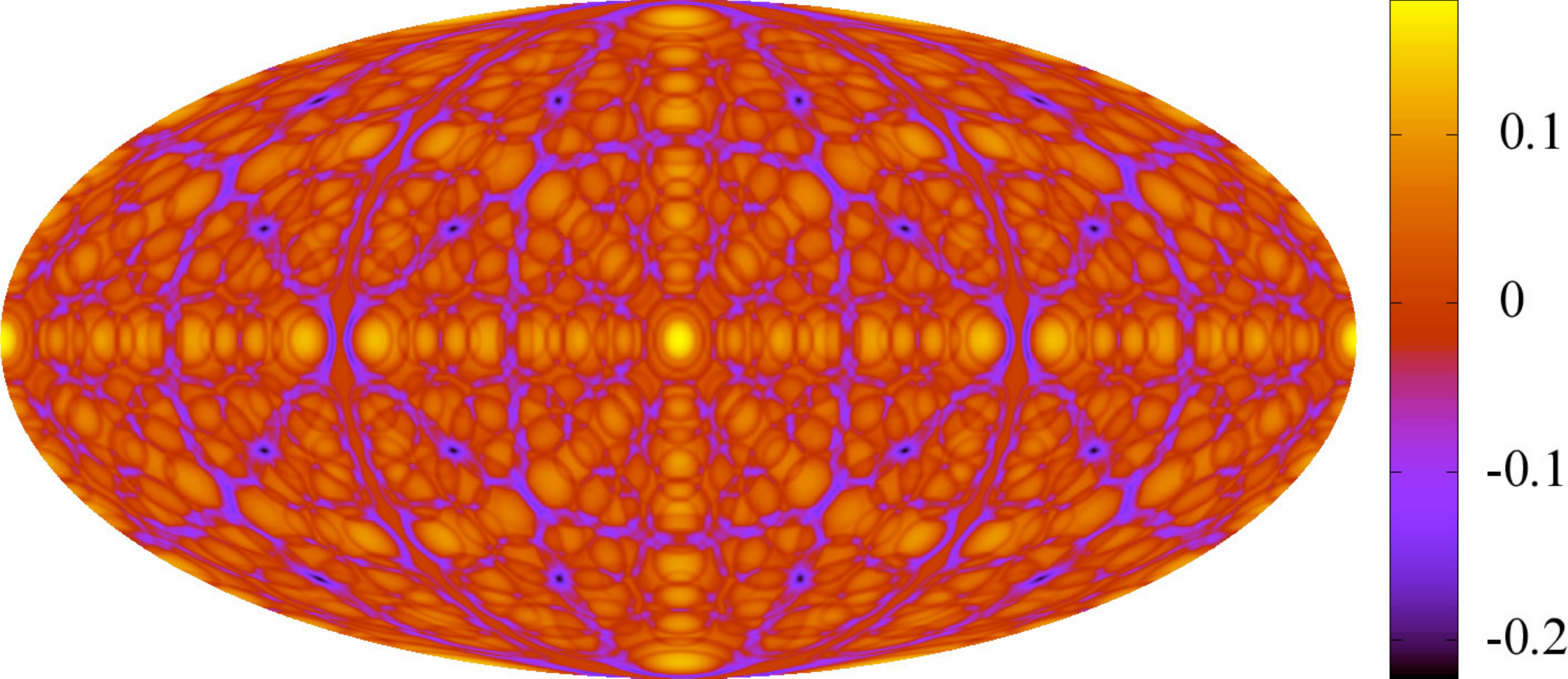}\\
\includegraphics[width=0.45\textwidth]{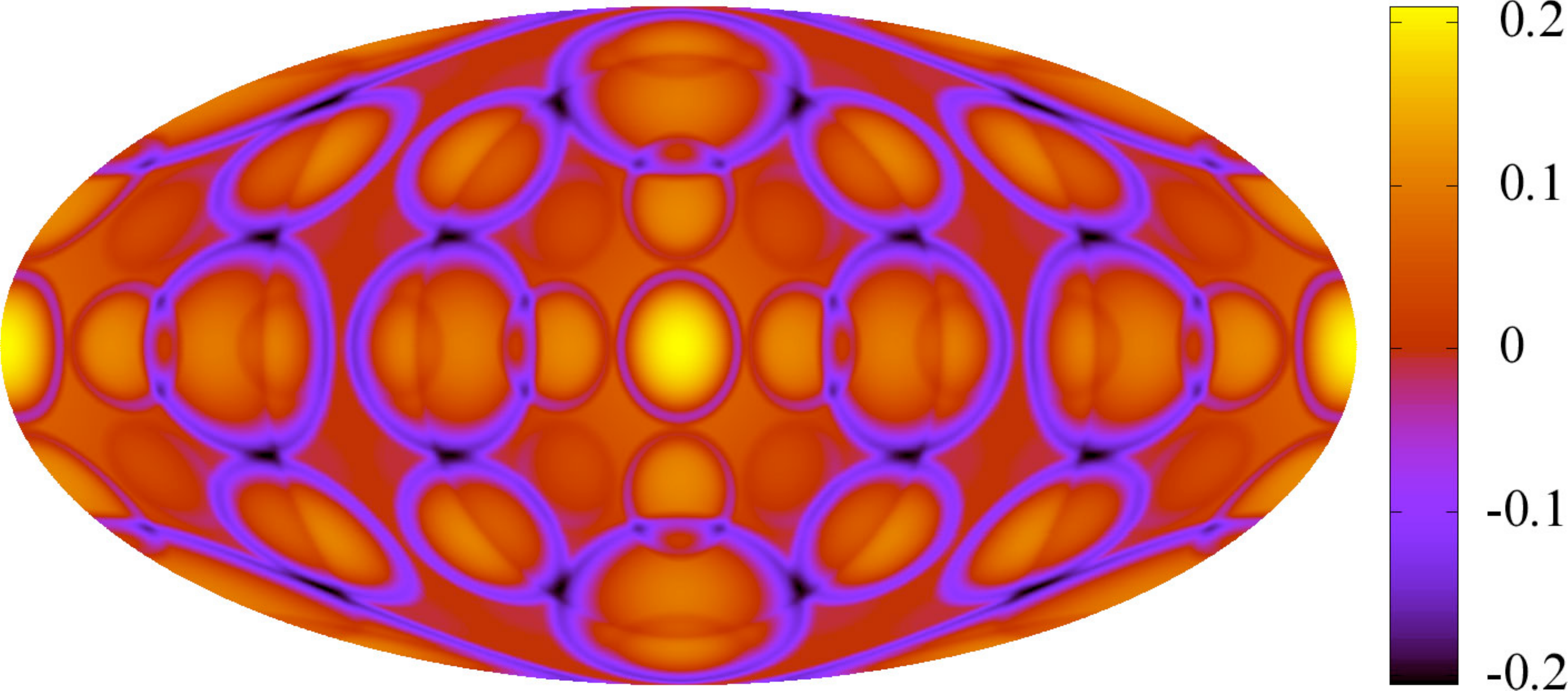}&
\includegraphics[width=0.45\textwidth]{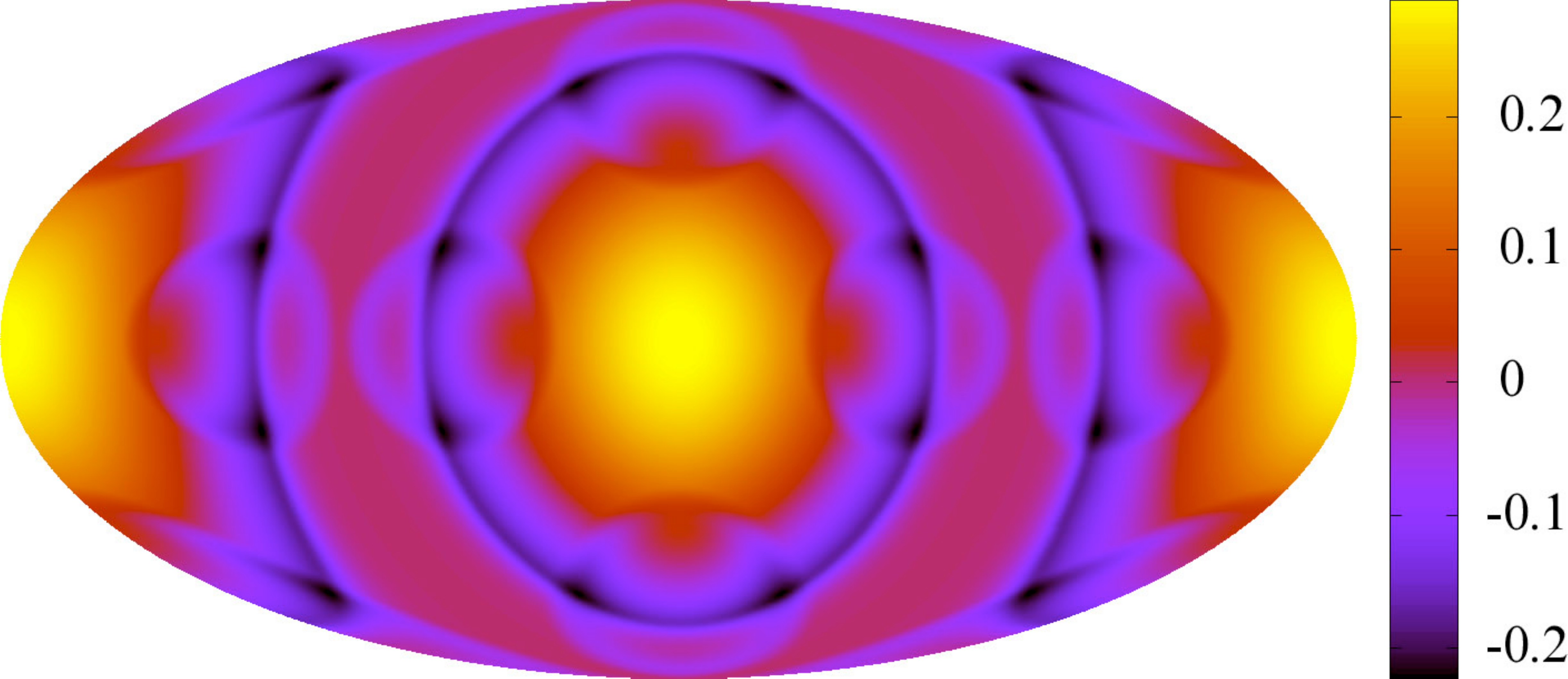}
\end{tabular}
\end{center}
\caption{The full-sky maps of the angular diameter distance at $z=1.92$ induced by the Swiss-Cheese structure. The quantity shown is $\frac{d_{\rm A}- \bar d_{\rm A}}{\bar d_{\rm A}}$. Again these are maps in Swiss-Cheese universes with holes of size $r_{\rm hl} =$ 3.5 Mpc (upper left figure), $r_{\rm hl} =$ 350 Mpc  (upper right figure), $r_{\rm hl} =$ 875 Mpc  (lower left figure) and $r_{\rm hl} =$ 1.75 Gpc  (lower right figure). In contrast with the redshift maps, in the angular diameter distance, the {\em maximum} anisotropy does not depend on the size of the holes.}\label{fig:damap}
\end{figure}

In figure~\ref{fig:damap} we show a full sky map of the angular diameter distance at a fixed redshift, namely the maximum redshift found in the temperature map of each cosmology at $t=\bar t$, which for each cosmology is close to $z = 1.92$. This means that in most directions the redshift of a photon was slightly less at time $t =\bar t$, hence the integration was continued through cheese only up to the right redshift. 
Interestingly, the maximum anisotropy in angular diameter distance is the same for all models, which was illustrated in the left of figure~\ref{fig:AnDL}.

\begin{figure}
\begin{center}
\includegraphics[angle=-90,width=.7\textwidth]{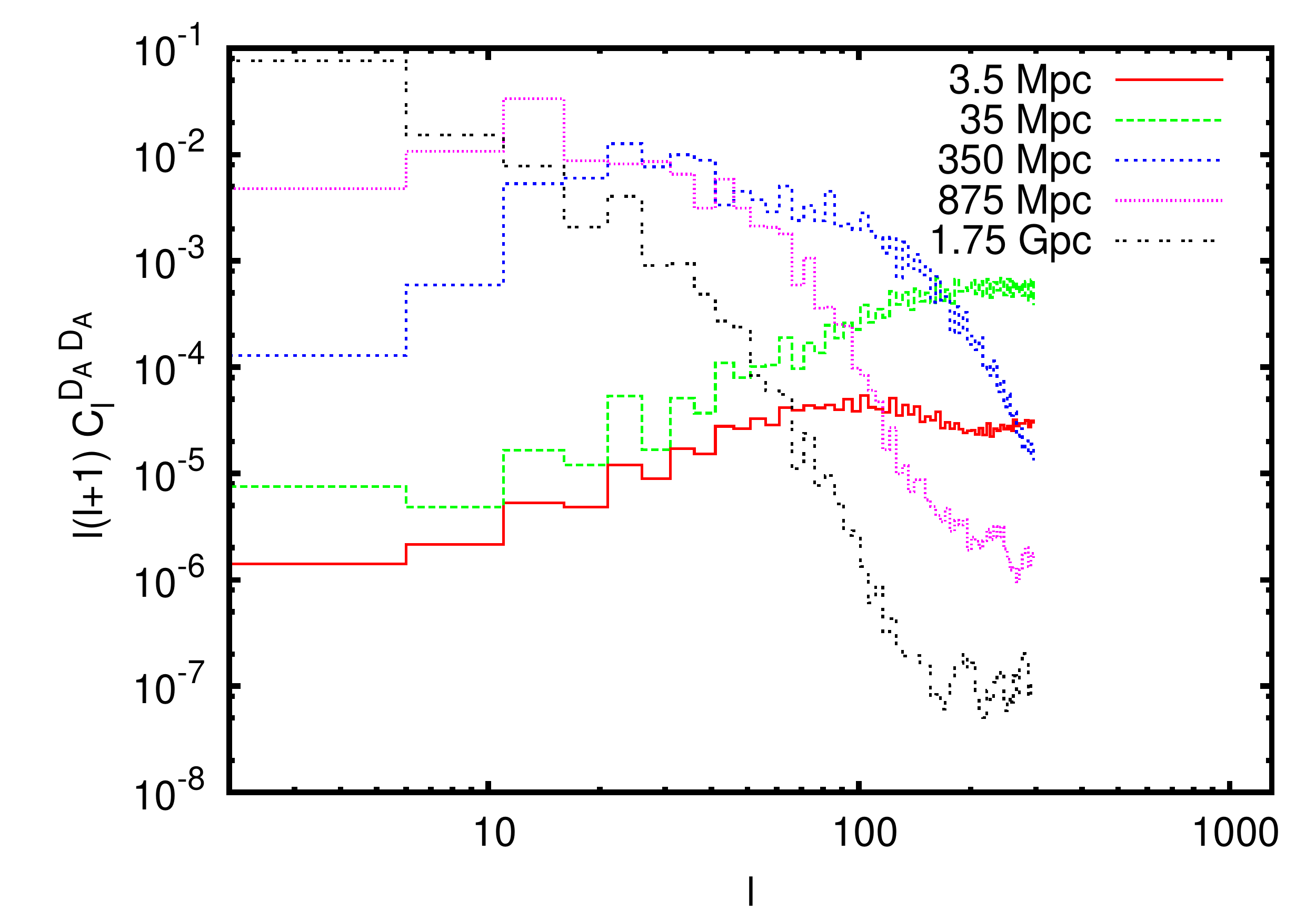}
\end{center}
\caption{The $C^{d_{\rm A} d_{\rm A}}_l$'s for different cosmologies, in bins of 5 multipoles.  Even though the maximum observed $d_{\rm A}$ is independent of the size of holes, the average anisotropy goes down with decreasing hole size. This is because the number of special directions decreases with decreasing hole size.}\label{fig:da-cls1}
\end{figure}
\begin{figure}
\vspace{-0.4cm}
\begin{center}
\begin{tabular}{rl}
\includegraphics[angle=-90,width=.45\textwidth]{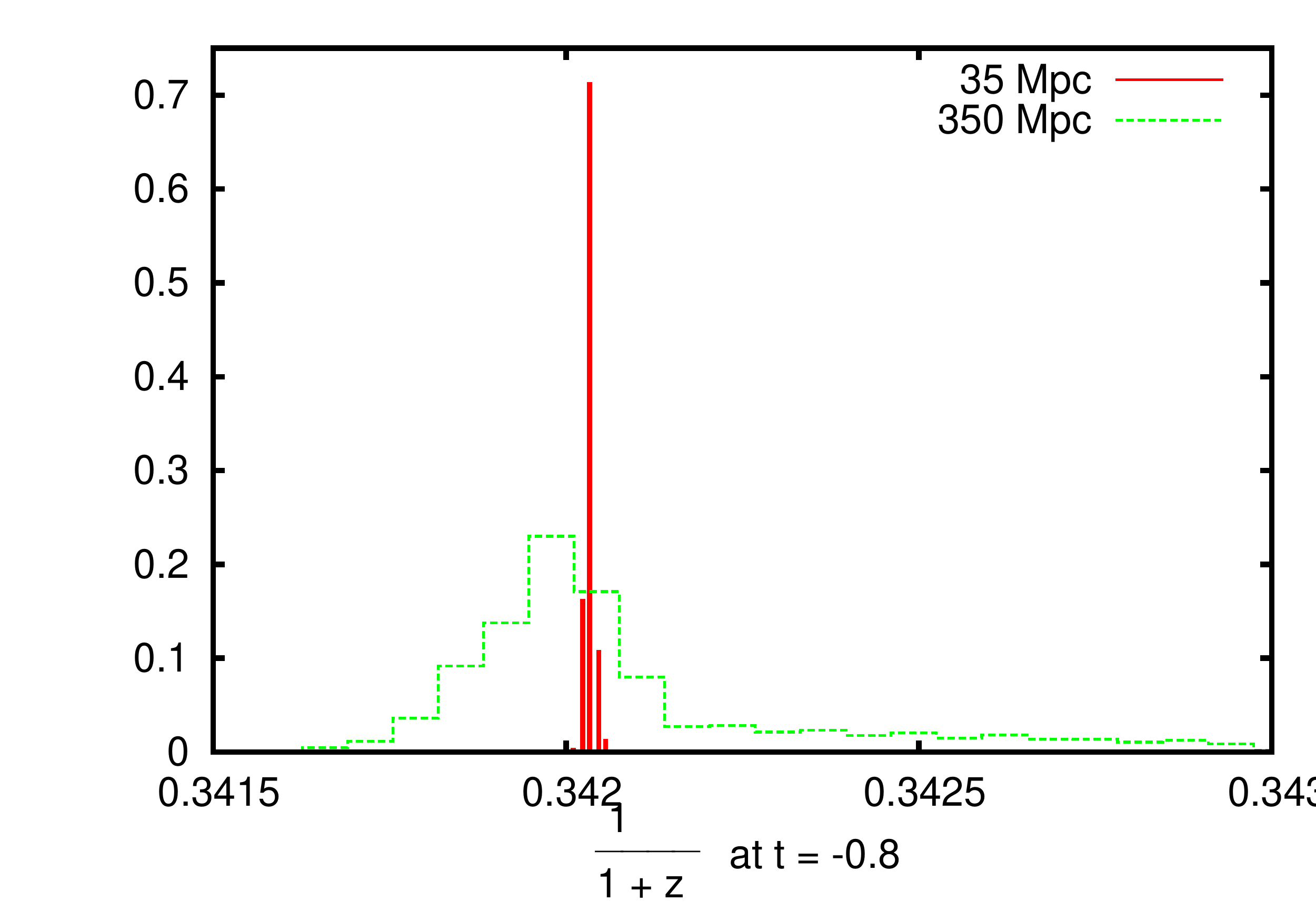}&
\includegraphics[angle=-90,width=.45\textwidth]{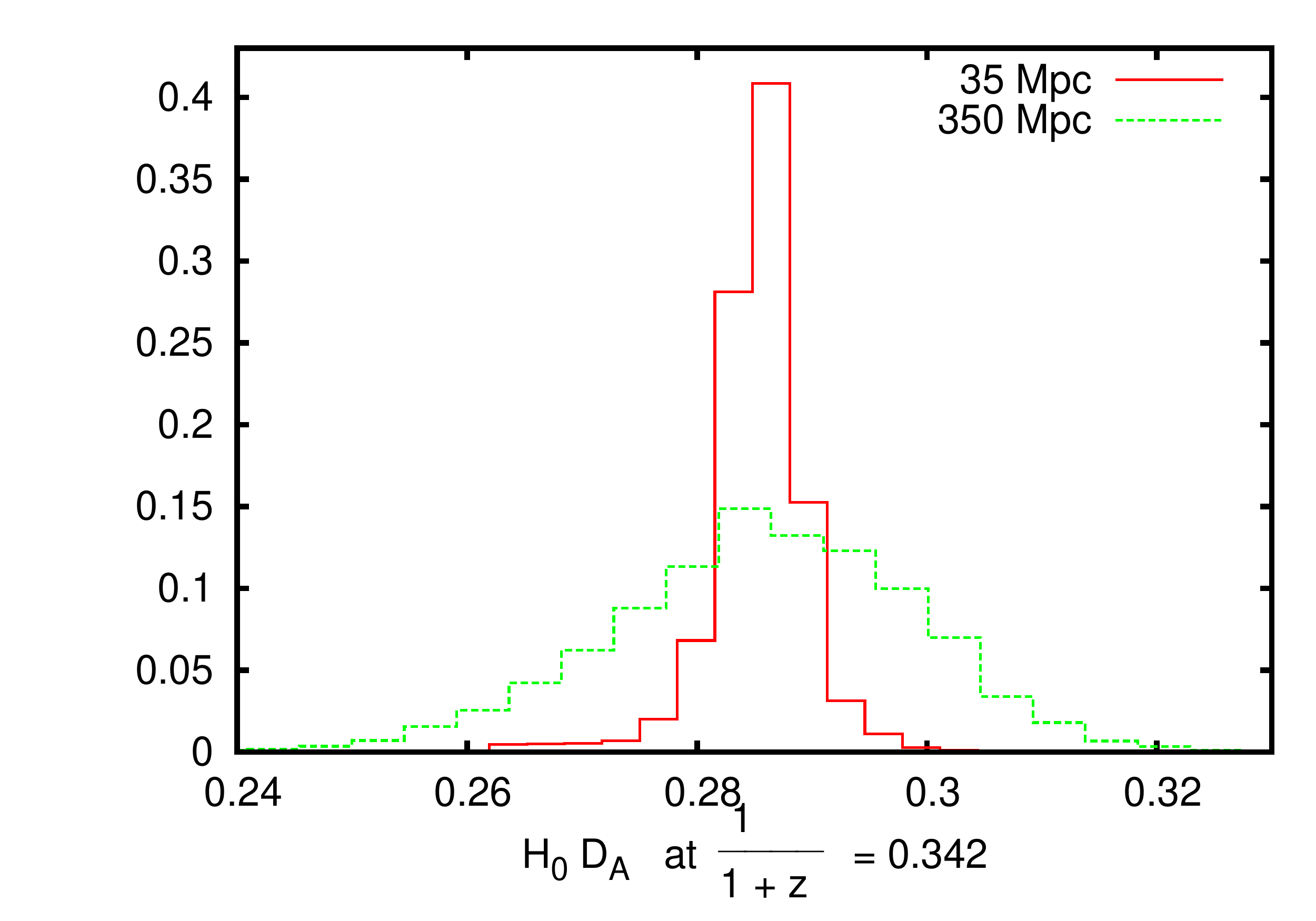}
\end{tabular}
\end{center}
\caption{{\em Left:} The distribution of redshift of photons in all directions, at fixed time. Clearly, for smaller holes, the variance in redshift decreases. {\em Right:} The distribution of angular diameter distance of photons in all directions, at fixed redshift. Again, for smaller holes, the variance decreases.}\label{fig:dists}
\end{figure}

In Ref.~\cite{Vanderveld:2008vi} it was already shown that, for holes of radius 350 Mpc, the average $d_{\rm A}$ corresponds to that of a cheese-only universe.
In figure \ref{fig:da-cls1} we show the angular power spectrum of the autocorrelation in the angular diameter map, $C_l^{\rm d_{\rm A}d_{\rm A}}$. This figure illustrates that, even though the maximum effect of the Swiss Cheese on $d_{\rm A}$ is about the same for all hole sizes, the number of special directions in which there is a high effect (or anti-effect), decreases with decreasing hole size. The smaller the size of holes is, the smaller the standard deviations in $d_A$ will be. This is also apparent in figure~\ref{fig:dists}, where the distributions of redshift $z$ and the angular diameter $d_{\rm A}$ are compared for Swiss-Cheese universes with two different hole sizes. The standard deviation in redshift goes down from $\sigma_{\frac{1}{1+z}}=2.6 \times 10^{-4}$ for $r_{\rm hl} =$ 350 Mpc to $\sigma_{\frac{1}{1+z}}=3.8 \times 10^{-5}$ for $r_{\rm hl} =$ 35 Mpc.
Similarly for the angular diameter distance, it goes down from $\sigma_{d_{\rm A}}=1.3 \times 10^{-2}$ to $\sigma_{d_{\rm A}}=4.3 \times 10^{-3}$.

\subsection{Dependence on the size of holes}

In Refs.~\cite{Biswas:2007gi,Brouzakis:2008uw} it was found that the effect of one hole on $d_{\rm A}$, with respect to a cheese-only passage, goes as $\Delta d_{\rm A}\equiv d^{\rm hole}_{\rm A} - d^{\rm cheese}_{\rm A}\sim r_{\rm hl}^3/R_{\rm H}^3$, with $R_{\rm H}$ the Hubble radius. Increasing the number of holes with $N_{\rm hl} \sim R_{\rm H}/r_{\rm hl}$, this would mean that the maximum effect scales as $d_{\rm A, max}\sim  r_{\rm hl}^2/R_{\rm H}^2$. 
As total lensing effect is crucial to the success of a Swiss-Cheese model, this contradictions deserves more attention. The difference in size-dependence may be an artefact of the modeling. However we find a difference at the analytical level, before even deciding which model to take. Let us explain why we come to a different conclusion.

\setlength{\totfigwidth}{\textwidth}
\begin{figure}
\setlength{\unitlength}{0.001\totfigwidth}
\begin{picture}(0,0)(0,-48)%
\includegraphics[width=.4\totfigwidth]{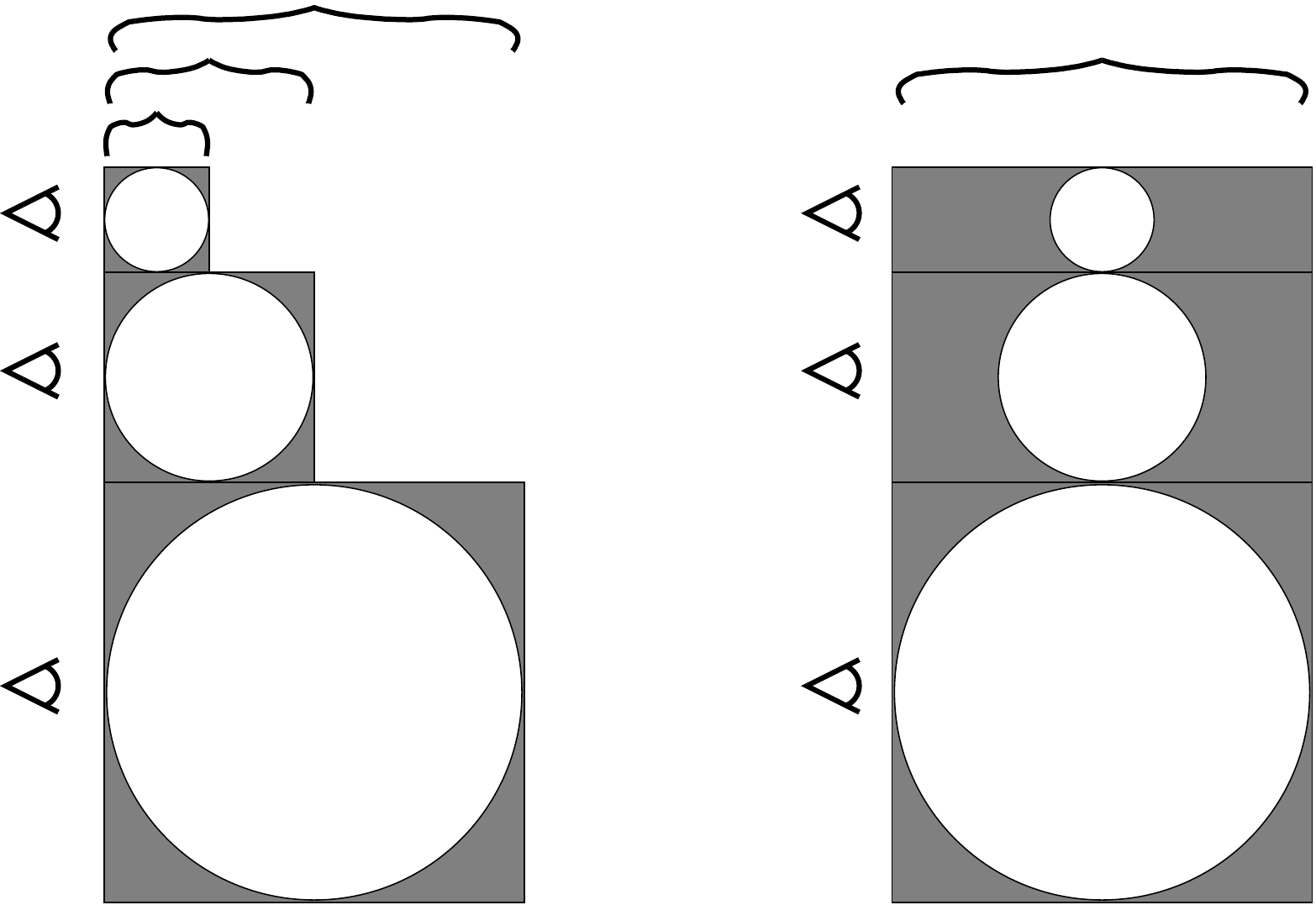}%
\end{picture}%
\begin{picture}(0,0)(-480,-335)%
\includegraphics[angle=-90,width=.5\totfigwidth]{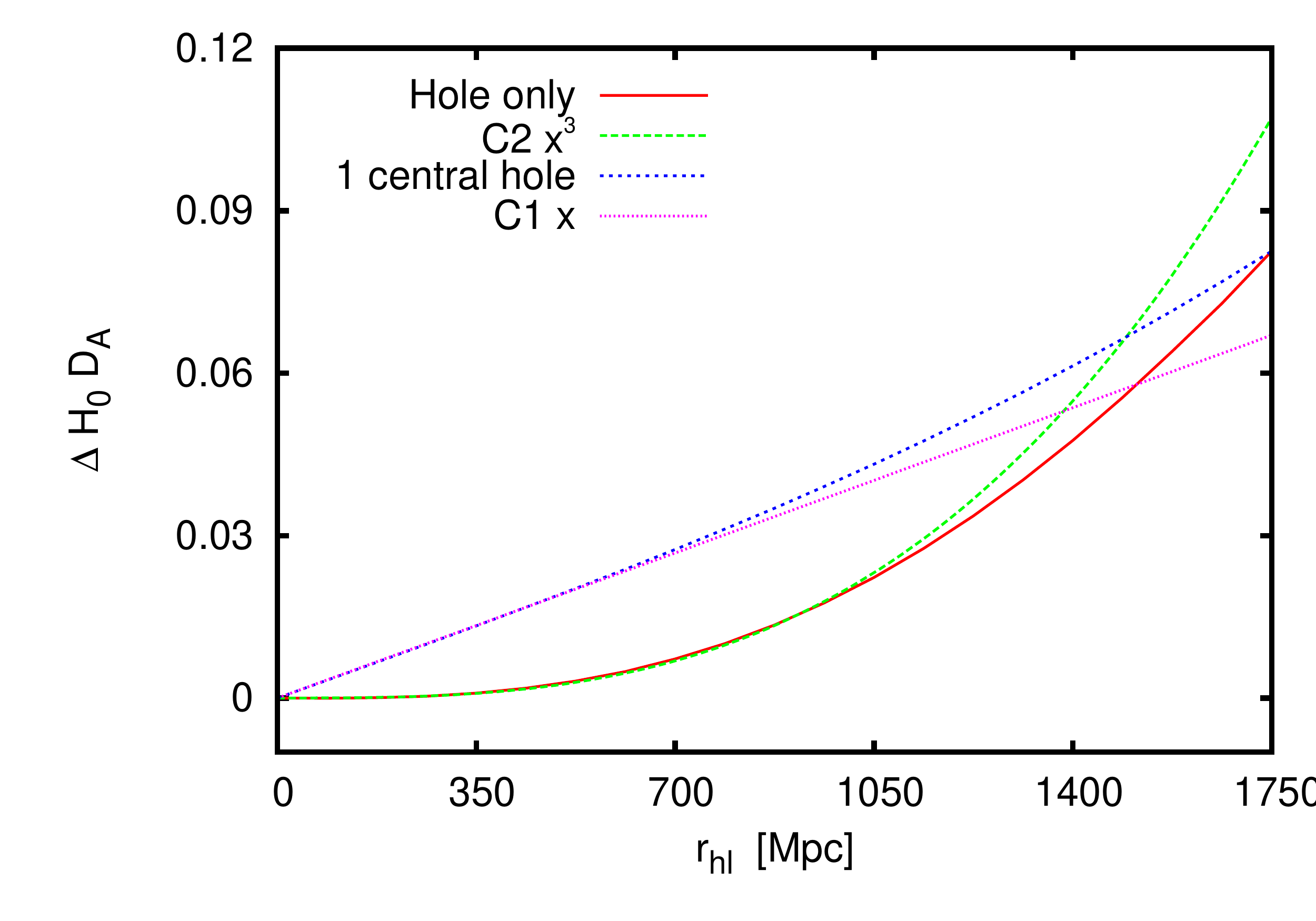}%
\end{picture}%
\begin{picture}(1000,466.15)(0,0)
\put(70,330){$d_{\rm A}(r)$}
\put(300,330){$d_{\rm A}(r)$}
\end{picture}%
\caption{
An ambiguity when speaking of the effect of the size of holes on $d_{\rm A}$. {\em Left:} Two sketches of methods comparing hole sizes. The first method considers only distance traveled through a hole, and compares $d_{\rm A}$ at the exit of that hole to that of an cheese-only photon. The second method considers one central void embedded in cheese, and compares $d_{\rm A}$ at the end of the same total comoving distance through a hole and cheese. {\em Right:} Using the first comparison method we recover the behaviour that is found in Refs.~\cite{Biswas:2007gi,Brouzakis:2008uw}, $\Delta d_{\rm A}\propto r_{\rm hl}^3$ (red, solid). Using the second method, we find a different behaviour which is closer to $\Delta d_{\rm A}\propto r_{\rm hl}$ (blue, dashed).
}\label{fig:delta_da_types}
\end{figure}
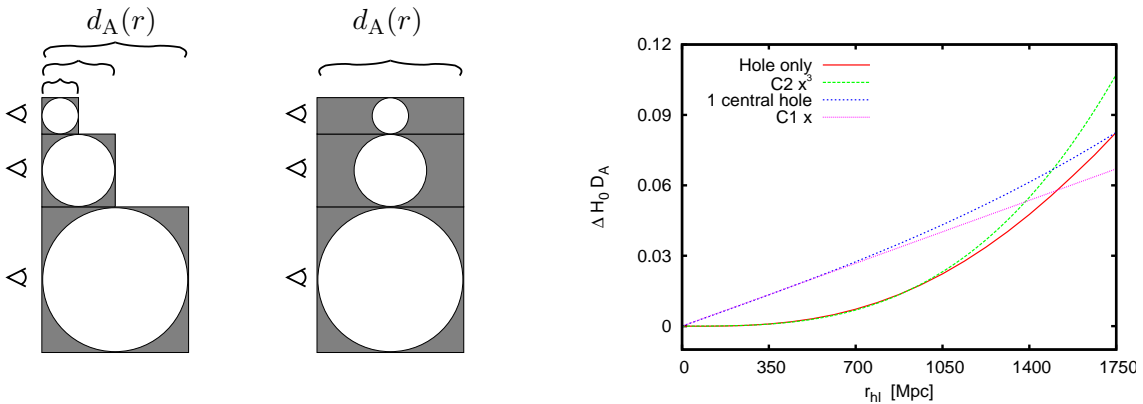

The calculations in Refs.~\cite{Biswas:2007gi,Brouzakis:2008uw} consider one hole at a time, and one hole only. The effect of several holes does however not grow linearly with the number of holes. It may happen to do so for a particular modeling, but in general it will not. In figure~\ref{fig:delta_da_types} we show the results of two different approaches. The first approach considers only one hole. If one starts an integration when entering a hole, and stops when exiting the hole in order to compare $d_{\rm A}$ with the value found when travelling the same distance through cheese, we recover in fact the foreseen effect of $\Delta d_{\rm A}\propto r_{\rm hl}^3$. In the second approach, we consider a photon passing one hole along a fixed distance, where all space outside the single hole is necessarily cheese. In this case we do not find the same behaviour, but the size dependende seems to be more like $\Delta d_{\rm A}\propto r_{\rm hl}$.

\begin{figure}
\begin{center}
\includegraphics[angle=-90,width=.47\textwidth]{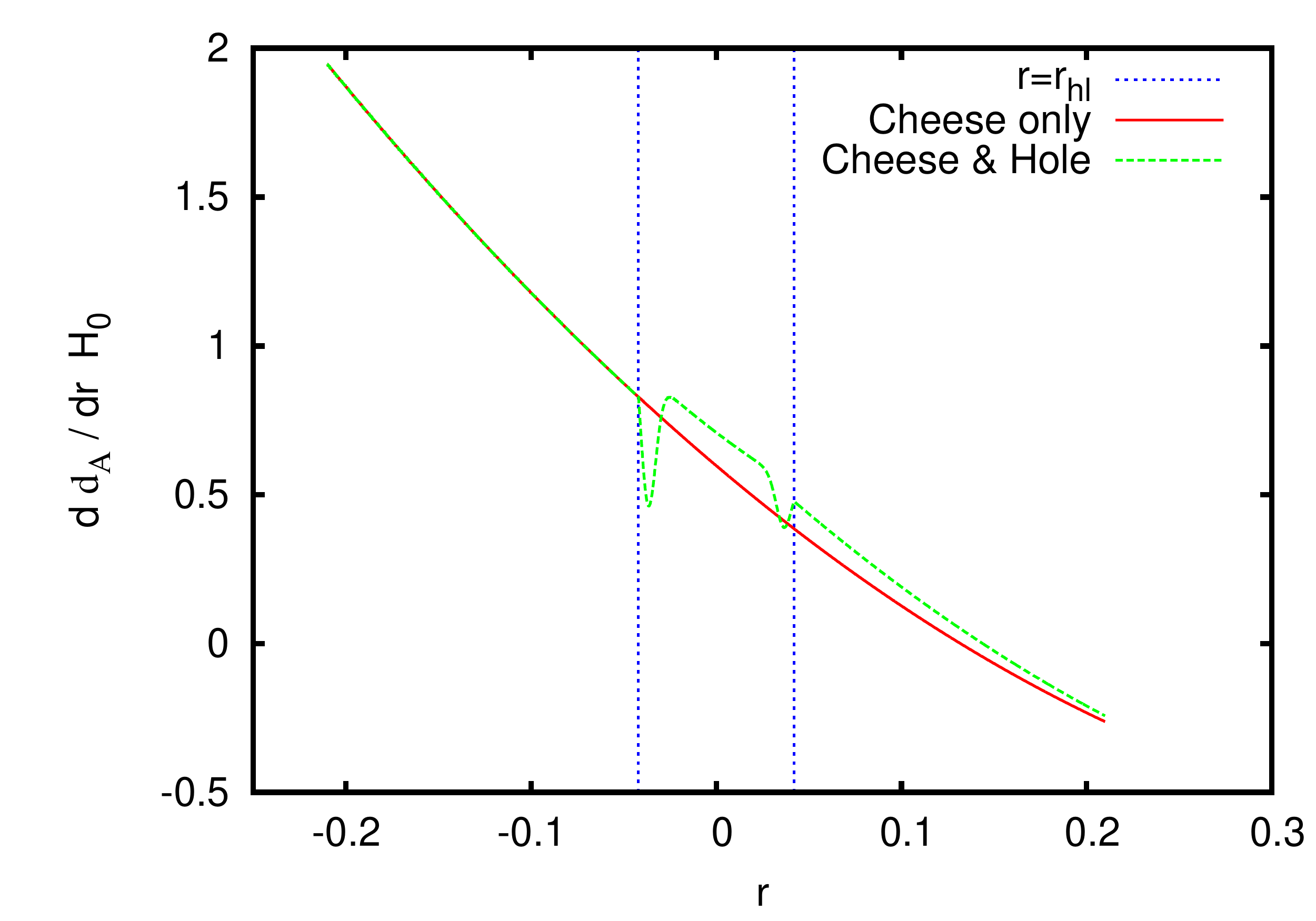}
\includegraphics[angle=-90,width=.47\textwidth]{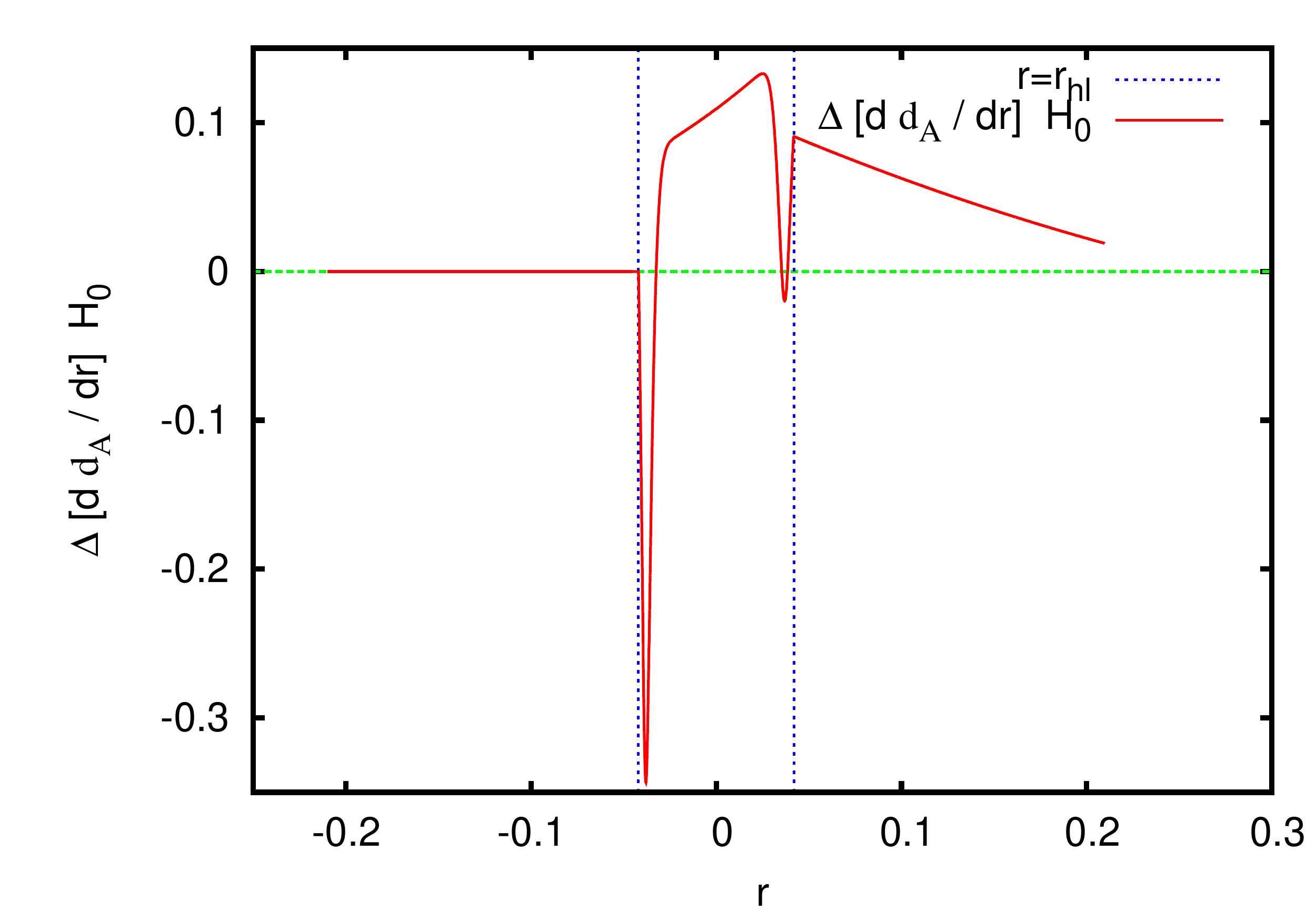}
\end{center}
\caption{{\em Left:} A comparison of the $d \,{\rm d}_{\rm A} / dr$ along the trajectories of two photons. Time increases from right to left, and the initial conditions are chosen such that an observer today (at $r=-0.21$) sees a source through cheese only (red solid line) and a source through one central hole, with $r_{\rm hl} = 350$ Mpc, centered between the observer and $t=\bar t$ (green dashed line). Both sources are observed at the same solid angle, which translates to $d \,{\rm d}_{\rm A} / dr$ at the observer. They have however a different size, which translates to $d \,{\rm d}_{\rm A} / dr$ at the source. That means that the source at $r=0.21$ behind the hole must have a larger area in order to be observed under the same solid angle, and hence has a larger angular diameter distance. The key point in this figure is the difference in $d \,{\rm d}_{\rm A} / dr$ at both sides of the hole. {\em Right:} $\Delta [d \,{\rm d}_{\rm A} / dr]$ shows the suppressing effect the hole has on the derivative of the area of the beam, causing the source to be observed under smaller solid angle than a source of the same size seen through cheese only.
 Hence, once a photon passes a hole, the lensing effect of the hole continues playing a role beyond the hole, by having changed the initial conditions of the integration beyond the hole.}\label{fig:sizedepda}
\end{figure}

The reason for this discrepancy is that, after passing a void, the quantity $\frac{d\sqrt{A}}{dr}$ (see equations~(\ref{eq:inisqa},\ref{eq:exactda2})) keeps a deviation from the same quantity for a cheese-only photon, as shown in figure~\ref{fig:sizedepda}. Hence, if one were to consider the part of the integration, starting only beyond the patch, the initial conditions for the cheese-only photon and the photon that passed through the hole are different. The quantity $\frac{d\sqrt{A}}{d\lambda}$ can be interpreted as a measure of divergence / convergence of the beam. Changing this quantity necessarily changes the final beamsize, even when no more holes are on the way. There is no reason to believe that this quantity obtains no change from a hole, as `the magnification of the beam, is proportional to the integrated column density along the beam's path'~\cite{Vanderveld:2008vi}. The integration is performed over the past light cone of a photon. The photon spends more time in the void of a hole than it does in the shell of a hole, therefore there is no cancellation of the defocussing effect when leaving a hole.
In fact, a closer look at Eqs.~(\ref{eq:geo1}-\ref{eq:geo2},\ref{eq:exactda1},\ref{eq:exactda2}) for the special case of a radial photon, such that $c_{\phi} = \xi = 0$, shows us,
\begin{align}
\frac{d^2\sqrt{A}}{dr^2}&=-\frac{2}{3}\sqrt{A}\,\,\frac{Y'(r,t)^2\rho(r,t) }{W(r)^2}\nonumber\\
& + \frac{d\sqrt{A}}{dr} \left[ \frac{Y''(r,t)}{Y'(r,t)}-\frac{W'(r)}{W(r)}+\frac{ \dot Y'(r,t)}{W(r)} \right],\nonumber\\
&=\sqrt{A} \,\, F_1(r,t) +\frac{d\sqrt{A}}{dr} F_2(r,t).
\end{align}
Here we used,
\begin{align}
\frac{d^2}{dr^2}=\left(\frac{d\lambda}{dr}\right)^2\frac{d^2}{d\lambda^2} + \frac{d^2\lambda}{dr^2}\frac{d}{d\lambda},
\end{align}
in order to be able to unambiguously compare photons along different geodesics, with different afine parameters $\lambda$ but with the same coordinates $r$ and $t$.
Now let us compare two photons, that entered a patch with the same initial conditions, and let one of them see only cheese. At the exit of the hole, the difference in $\frac{d^2\sqrt{A}}{dr^2}$ is,
\begin{align}
\Delta \left[ \frac{d^2\sqrt{A}}{dr^2} \right]&=\Delta\left[ \sqrt{A} \right]\,\, F_1(r_{\rm hl},t) + \Delta \left[\frac{d\sqrt{A}}{dr}\right] F_2(r_{\rm hl},t)
\end{align}
where in we use the notation $\Delta[ Q ]= Q_{\rm hole\,\,photon}(r_{\rm hl}) - Q_{\rm cheese \,\,photon}(r_{\rm hl})$.

First consider $F_1(r_{\rm hl},t)$ and $F_2(r_{\rm hl},t)$. These are functions of background quantities only, evaluated at $r=r_{\rm hl}$ where all background quantities are supposed to continuously match from the hole to the cheese. Hence, these functions are the same for all photons, no matter whether they saw a hole or cheese only, no matter what initial conditions their beam had.
By construction we have $\Delta\left[ \sqrt{A} \right] \neq 0$, since that is the whole purpose of the Swiss-Cheese structure. Therefore, beyond the hole, one expects that at least either $ \Delta \left[ \frac{d\sqrt{A}}{dr}\right]\neq0$ or $\Delta \left[ \frac{d^2\sqrt{A}}{dr^2}\right]\neq0$ if $\Delta\left[ \sqrt{A}\right] \neq 0$, proving that the defocussing effect of the hole carries beyond the hole. 

The reader should keep in mind that the defocussing effect is of importance when one considers the maximum effect one hole can have on $d_{\rm A}$, even beyond the passage of that particular hole. Remember that we are considering the divergence of a beam, after having passed a hole, which is the derivative of the angular diameter distance. However, as it has been shown in Refs.~\cite{Vanderveld:2008vi} that the average effect of one hole on the angular diameter distance disappears when averaged over angles of incidence, it is likely that the average effect of one hole on the divergence of the beam (which, as we just discussed, influences the angular diameter distance beyond the hole) will also disappear on average.

A final remark to make here, is a reminder that we are dealing with a toy model. The model is constructed manually and ideally. The defocussing by the hole may be an artefact of the modeling, and may disappear in a more realistic model. We showed that in idealized models it is most likely that the divergence of a beam {\em is} altered by a hole, but this gives no guarantee for more realistic models.

\section{Discussion and conclusion}\label{sec:conclusion}
For the first time we have performed a full sky simulation of the CMB in Swiss-Cheese universes, in realizations with different hole sizes. We have shown that if all holes have a radius larger than $r_{\rm hl}=35$ Mpc, the Swiss-Cheese model can be ruled out on basis of the observed CMB. One could try to change the density profile of the holes, but in order to leave the CMB intact, that new profile would necessarily be closer to FLRW, hence no longer reproducing the wanted angular diameter distance effect. 

Another option to try to save the model is to decrease the size of holes. We find that for small holes, the maximum angular diameter effect remains the same as for larger holes. However, the probability that special directions exist, in which one sees through the centers of a series of perfectly aligned holes,
decreases with decreasing hole size, at least in the case of holes on a lattice, as in this work.

In Refs.~\cite{1976ApJ...208L...1W,Vanderveld:2008vi} it was argued that (spherical) mass compensated holes in a Swiss-Cheese universe cannot make the average angular diameter distance depart from the EdS-case, when the observer is outside the hole. Not surprisingly, we confirm this result. Moreover, we have shown that even though the average $d_{\rm A}$ does not change, the anisotropy in CMB-temperature due to the same hole does in fact differ significantly from the EdS-case, which in principle should not depend on the sphericalness of a hole, but on the mere fact that the observer sees both through and past the side of the hole, inevitably leaving a Rees-Sciama imprint on the CMB.

These findings together lead to the conclusion that the Swiss-Cheese universe, as considered here, with mass compensated holes, all of them larger than $r_{\rm hl}=35$ Mpc, spherical or not, is ruled out. Based on the findings in Ref.~\cite{Vanderveld:2008vi}, and the findings in this work with respect to the variance in $d_{\rm A}$ in different directions that decreases with decreasing hole size, also Swiss-Cheese universes with
{\em spherical} mass compensated holes of radius smaller than $r_{\rm hl} =$ 35 Mpc are ruled out.

This conclusion applies to all Swiss-Cheese-like models in which the universe is saturated with mass compensated voids, and that try to explain the angular diameter distance at high redshift. 
We did not consider the case where our local Hubble volume contains just a small number of voids, without being saturated with them. Such a scenario remains interesting in order to explain anomalies in the CMB, such as the cold spot~\cite{Masina:2008zv}. However, such a scenario is in principle unrelated to possible alternative explanations for the observed accelerated expansion.

We have shown that the maximum effect of holes on $d_{\rm A}$ has weaker, if not zero, dependence  on the size of the holes  than previously claimed in the literature, due to the defocussing effect a hole has, which carries influence beyond the hole. We assessed the importance of taking into account this defocussing, when addressing the effect of different hole sizes.
 This effect could be due to the modeling of the density profile. The model dependence should be more carefully assessed, before definitive conclusions can be drawn.
It remains to be investigated as well, what the effect would be of non-spherical holes, with a typical size $r_{\rm hl} < 35$ Mpc, densely spread in such a manner that locally the universe is nowhere described by the cheese metric, but globally remains FLRW. Another addition to subject to research, is to see the effect of virialization of the mass shells, which, as pointed out in Ref.~\cite{Biswas:2007gi}, may increase the effect on $d_{\rm A}$, and at the same time should decrease the Rees-Sciama effect, hence decreasing the potential trouble with the CMB.

\acknowledgments
The author is supported
by the EU 6th Framework Marie Curie Research and Training network
``UniverseNet'' (MRTN-CT-2006-035863). Numerical simulations were
performed on the MUST cluster at LAPP (CNRS \& Universit\'e de
Savoie).  It is a
pleasure to thank Julien Lesgourgues and Alessio Notari for helpful comments and discussions. In particular the author thanks Valerio Marra for fruitful discussions over an excellent risotto.

\bibliographystyle{JHEP}
\bibliography{refs}

\providecommand{\href}[2]{#2}\begingroup\raggedright\begin{thebibliography}{10}

\bibitem{Riess:1998cb}
{\bf Supernova Search Team} Collaboration, A.~G. Riess {\em et.~al.}, {\it
  {Observational Evidence from Supernovae for an Accelerating Universe and a
  Cosmological Constant}},  {\em Astron. J.} {\bf 116} (1998) 1009--1038,
  [\href{http://xxx.lanl.gov/abs/astro-ph/9805201}{{\tt astro-ph/9805201}}].

\bibitem{Astier:2005qq}
{\bf The SNLS} Collaboration, P.~Astier {\em et.~al.}, {\it {The Supernova
  Legacy Survey: Measurement of $\Omega_M$, $\Omega_\Lambda$ and w from the
  First Year Data Set}},  {\em Astron. Astrophys.} {\bf 447} (2006) 31--48,
  [\href{http://xxx.lanl.gov/abs/astro-ph/0510447}{{\tt astro-ph/0510447}}].

\bibitem{Freedman:2000cf}
{\bf HST} Collaboration, W.~L. Freedman {\em et.~al.}, {\it {Final Results from
  the Hubble Space Telescope Key Project to Measure the Hubble Constant}},
  {\em Astrophys. J.} {\bf 553} (2001) 47--72,
  [\href{http://xxx.lanl.gov/abs/astro-ph/0012376}{{\tt astro-ph/0012376}}].

\bibitem{Smoot:1992td}
G.~F. Smoot {\em et.~al.}, {\it {Structure in the COBE differential microwave
  radiometer first year maps}},  {\em Astrophys. J.} {\bf 396} (1992) L1--L5.

\bibitem{Kuo:2002ua}
{\bf ACBAR} Collaboration, C.-l. Kuo {\em et.~al.}, {\it {High Resolution
  Observations of the CMB Power Spectrum with ACBAR}},  {\em Astrophys. J.}
  {\bf 600} (2004) 32--51,
  [\href{http://xxx.lanl.gov/abs/astro-ph/0212289}{{\tt astro-ph/0212289}}].

\bibitem{Mason:2002tm}
B.~S. Mason {\em et.~al.}, {\it {The Anisotropy of the Microwave Background to
  l = 3500: Deep Field Observations with the Cosmic Background Imager}},  {\em
  Astrophys. J.} {\bf 591} (2003) 540--555,
  [\href{http://xxx.lanl.gov/abs/astro-ph/0205384}{{\tt astro-ph/0205384}}].

\bibitem{Komatsu:2008hk}
{\bf WMAP} Collaboration, E.~Komatsu {\em et.~al.}, {\it {Five-Year Wilkinson
  Microwave Anisotropy Probe Observations:Cosmological Interpretation}},  {\em
  Astrophys. J. Suppl.} {\bf 180} (2009) 330--376,
  [\href{http://xxx.lanl.gov/abs/0803.0547}{{\tt arXiv:0803.0547}}].

\bibitem{Rasanen:2008be}
S.~Rasanen, {\it {Light propagation in statistically homogeneous and isotropic
  dust universes}},  {\em JCAP} {\bf 0902} (2009) 011,
  [\href{http://xxx.lanl.gov/abs/0812.2872}{{\tt arXiv:0812.2872}}].

\bibitem{Efstathiou:2001cw}
{\bf 2dFGRS} Collaboration, G.~Efstathiou {\em et.~al.}, {\it {Evidence for a
  non-zero Lambda and a low matter density from a combined analysis of the 2dF
  Galaxy Redshift Survey and Cosmic Microwave Background Anisotropies}},  {\em
  Mon. Not. Roy. Astron. Soc.} {\bf 330} (2002) L29,
  [\href{http://xxx.lanl.gov/abs/astro-ph/0109152}{{\tt astro-ph/0109152}}].

\bibitem{Eisenstein:2005su}
{\bf SDSS} Collaboration, D.~J. Eisenstein {\em et.~al.}, {\it {Detection of
  the Baryon Acoustic Peak in the Large-Scale Correlation Function of SDSS
  Luminous Red Galaxies}},  {\em Astrophys. J.} {\bf 633} (2005) 560--574,
  [\href{http://xxx.lanl.gov/abs/astro-ph/0501171}{{\tt astro-ph/0501171}}].

\bibitem{Percival:2007yw}
W.~J. Percival {\em et.~al.}, {\it {Measuring the Baryon Acoustic Oscillation
  scale using the SDSS and 2dFGRS}},  {\em Mon. Not. Roy. Astron. Soc.} {\bf
  381} (2007) 1053--1066, [\href{http://xxx.lanl.gov/abs/0705.3323}{{\tt
  arXiv:0705.3323}}].

\bibitem{Seikel:2007pk}
M.~Seikel and D.~J. Schwarz, {\it {How strong is the evidence for accelerated
  expansion?}},  {\em JCAP} {\bf 0802} (2008) 007,
  [\href{http://xxx.lanl.gov/abs/0711.3180}{{\tt arXiv:0711.3180}}].

\bibitem{Celerier:1999hp}
M.-N. Celerier, {\it {Do we really see a cosmological constant in the
  supernovae data ?}},  {\em Astron. Astrophys.} {\bf 353} (2000) 63--71,
  [\href{http://xxx.lanl.gov/abs/astro-ph/9907206}{{\tt astro-ph/9907206}}].

\bibitem{Tomita:2000rf}
K.~Tomita, {\it {Anisotropy of the Hubble Constant in a Cosmological Model with
  a Local Void on Scales of ~ 200 Mpc}},  {\em Prog. Theor. Phys.} {\bf 105}
  (2001) 419, [\href{http://xxx.lanl.gov/abs/astro-ph/0005031}{{\tt
  astro-ph/0005031}}].

\bibitem{Tomita:2000jj}
K.~Tomita, {\it {A Local Void and the Accelerating Universe}},  {\em Mon. Not.
  Roy. Astron. Soc.} {\bf 326} (2001) 287,
  [\href{http://xxx.lanl.gov/abs/astro-ph/0011484}{{\tt astro-ph/0011484}}].

\bibitem{Tomita:2001gh}
K.~Tomita, {\it {Analyses of Type Ia Supernova Data in Cosmological Models with
  a Local Void}},  {\em Prog. Theor. Phys.} {\bf 106} (2001) 929--939,
  [\href{http://xxx.lanl.gov/abs/astro-ph/0104141}{{\tt astro-ph/0104141}}].

\bibitem{Iguchi:2001sq}
H.~Iguchi, T.~Nakamura, and K.-i. Nakao, {\it {Is dark energy the only solution
  to the apparent acceleration of the present universe?}},  {\em Prog. Theor.
  Phys.} {\bf 108} (2002) 809--818,
  [\href{http://xxx.lanl.gov/abs/astro-ph/0112419}{{\tt astro-ph/0112419}}].

\bibitem{Tomita:2002df}
K.~Tomita, {\it {Dipole anisotropies of IRAS galaxies and the contribution of a
  large-scale local void}},  {\em Astrophys. J.} {\bf 584} (2003) 580--584,
  [\href{http://xxx.lanl.gov/abs/astro-ph/0211137}{{\tt astro-ph/0211137}}].

\bibitem{Moffat:2005yx}
J.~W. Moffat, {\it {Cosmic Microwave Background, Accelerating Universe and
  Inhomogeneous Cosmology}},  {\em JCAP} {\bf 0510} (2005) 012,
  [\href{http://xxx.lanl.gov/abs/astro-ph/0502110}{{\tt astro-ph/0502110}}].

\bibitem{Moffat:2005zx}
J.~W. Moffat, {\it {Large scale cosmological inhomogeneities, inflation and
  acceleration without dark energy}},
  \href{http://xxx.lanl.gov/abs/astro-ph/0504004}{{\tt astro-ph/0504004}}.

\bibitem{Moffat:2005ii}
J.~W. Moffat, {\it {Late-time inhomogeneity and acceleration without dark
  energy}},  {\em JCAP} {\bf 0605} (2006) 001,
  [\href{http://xxx.lanl.gov/abs/astro-ph/0505326}{{\tt astro-ph/0505326}}].

\bibitem{Alnes:2005rw}
H.~Alnes, M.~Amarzguioui, and O.~Gron, {\it {An inhomogeneous alternative to
  dark energy?}},  {\em Phys. Rev.} {\bf D73} (2006) 083519,
  [\href{http://xxx.lanl.gov/abs/astro-ph/0512006}{{\tt astro-ph/0512006}}].

\bibitem{Mansouri:2005rf}
R.~Mansouri, {\it {Structured FRW universe leads to acceleration: A non-
  perturbative approach}},
  \href{http://xxx.lanl.gov/abs/astro-ph/0512605}{{\tt astro-ph/0512605}}.

\bibitem{Vanderveld:2006rb}
R.~A. Vanderveld, E.~E. Flanagan, and I.~Wasserman, {\it {Mimicking Dark Energy
  with Lemaitre-Tolman-Bondi Models: Weak Central Singularities and Critical
  Points}},  {\em Phys. Rev.} {\bf D74} (2006) 023506,
  [\href{http://xxx.lanl.gov/abs/astro-ph/0602476}{{\tt astro-ph/0602476}}].

\bibitem{Garfinkle:2006sb}
D.~Garfinkle, {\it {Inhomogeneous spacetimes as a dark energy model}},  {\em
  Class. Quant. Grav.} {\bf 23} (2006) 4811--4818,
  [\href{http://xxx.lanl.gov/abs/gr-qc/0605088}{{\tt gr-qc/0605088}}].

\bibitem{Biswas:2006ub}
T.~Biswas, R.~Mansouri, and A.~Notari, {\it {Nonlinear Structure Formation and
  Apparent Acceleration: an Investigation}},  {\em JCAP} {\bf 0712} (2007) 017,
  [\href{http://xxx.lanl.gov/abs/astro-ph/0606703}{{\tt astro-ph/0606703}}].

\bibitem{Chung:2006xh}
D.~J.~H. Chung and A.~E. Romano, {\it {Mapping Luminosity-Redshift Relationship
  to LTB Cosmology}},  {\em Phys. Rev.} {\bf D74} (2006) 103507,
  [\href{http://xxx.lanl.gov/abs/astro-ph/0608403}{{\tt astro-ph/0608403}}].

\bibitem{Alnes:2006uk}
H.~Alnes and M.~Amarzguioui, {\it {The supernova Hubble diagram for off-center
  observers in a spherically symmetric inhomogeneous universe}},  {\em Phys.
  Rev.} {\bf D75} (2007) 023506,
  [\href{http://xxx.lanl.gov/abs/astro-ph/0610331}{{\tt astro-ph/0610331}}].

\bibitem{Caldwell:2007yu}
R.~R. Caldwell and A.~Stebbins, {\it {A Test of the Copernican Principle}},
  {\em Phys. Rev. Lett.} {\bf 100} (2008) 191302,
  [\href{http://xxx.lanl.gov/abs/0711.3459}{{\tt arXiv:0711.3459}}].

\bibitem{Alexander:2007xx}
S.~Alexander, T.~Biswas, A.~Notari, and D.~Vaid, {\it {Local Void vs Dark
  Energy: Confrontation with WMAP and Type Ia Supernovae}},
  \href{http://xxx.lanl.gov/abs/0712.0370}{{\tt arXiv:0712.0370}}.

\bibitem{GarciaBellido:2008nz}
J.~Garcia-Bellido and T.~Haugboelle, {\it {Confronting Lemaitre-Tolman-Bondi
  models with Observational Cosmology}},  {\em JCAP} {\bf 0804} (2008) 003,
  [\href{http://xxx.lanl.gov/abs/0802.1523}{{\tt arXiv:0802.1523}}].

\bibitem{Clifton:2008hv}
T.~Clifton, P.~G. Ferreira, and K.~Land, {\it {Living in a Void: Testing the
  Copernican Principle with Distant Supernovae}},  {\em Phys. Rev. Lett.} {\bf
  101} (2008) 131302, [\href{http://xxx.lanl.gov/abs/0807.1443}{{\tt
  arXiv:0807.1443}}].

\bibitem{GarciaBellido:2008gd}
J.~Garcia-Bellido and T.~Haugboelle, {\it {Looking the void in the eyes - the
  kSZ effect in LTB models}},  {\em JCAP} {\bf 0809} (2008) 016,
  [\href{http://xxx.lanl.gov/abs/0807.1326}{{\tt arXiv:0807.1326}}].

\bibitem{Hunt:2008wp}
P.~Hunt and S.~Sarkar, {\it {Constraints on large scale voids from WMAP-5 and
  SDSS}},  \href{http://xxx.lanl.gov/abs/0807.4508}{{\tt arXiv:0807.4508}}.

\bibitem{Krauss:1995yb}
L.~M. Krauss and M.~S. Turner, {\it {The cosmological constant is back}},  {\em
  Gen. Rel. Grav.} {\bf 27} (1995) 1137--1144,
  [\href{http://xxx.lanl.gov/abs/astro-ph/9504003}{{\tt astro-ph/9504003}}].

\bibitem{Biswas:2007gi}
T.~Biswas and A.~Notari, {\it {Swiss-Cheese Inhomogeneous Cosmology \& the Dark
  Energy Problem}},  {\em JCAP} {\bf 0806} (2008) 021,
  [\href{http://xxx.lanl.gov/abs/astro-ph/0702555}{{\tt astro-ph/0702555}}].

\bibitem{Kolb:2008bn}
E.~W. Kolb, V.~Marra, and S.~Matarrese, {\it {On the description of our
  cosmological spacetime as a perturbed conformal Newtonian metric and
  implications for the backreaction proposal for the accelerating universe}},
  {\em Phys. Rev.} {\bf D78} (2008) 103002,
  [\href{http://xxx.lanl.gov/abs/0807.0401}{{\tt arXiv:0807.0401}}].

\bibitem{Ellis:2005uz}
G.~F.~R. Ellis and T.~Buchert, {\it {The universe seen at different scales}},
  {\em Phys. Lett.} {\bf A347} (2005) 38--46,
  [\href{http://xxx.lanl.gov/abs/gr-qc/0506106}{{\tt gr-qc/0506106}}].

\bibitem{Rasanen:2006kp}
S.~Rasanen, {\it {Accelerated expansion from structure formation}},  {\em JCAP}
  {\bf 0611} (2006) 003, [\href{http://xxx.lanl.gov/abs/astro-ph/0607626}{{\tt
  astro-ph/0607626}}].

\bibitem{Buchert:2007ik}
T.~Buchert, {\it {Dark Energy from Structure - A Status Report}},  {\em Gen.
  Rel. Grav.} {\bf 40} (2008) 467--527,
  [\href{http://xxx.lanl.gov/abs/0707.2153}{{\tt arXiv:0707.2153}}].

\bibitem{Kolb:2009rp}
E.~W. Kolb, V.~Marra, and S.~Matarrese, {\it {Cosmological background solutions
  and cosmological backreactions}},
  \href{http://xxx.lanl.gov/abs/0901.4566}{{\tt arXiv:0901.4566}}.

\bibitem{Kundt:2009ky}
W.~Kundt, {\it {Critical Thoughts on Cosmology}},
  \href{http://xxx.lanl.gov/abs/0902.3151}{{\tt arXiv:0902.3151}}.

\bibitem{1969ApJ...155...89K}
R.~{Kantowski}, {\it {Corrections in the Luminosity-Redshift Relations of the
  Homogeneous Fried-Mann Models}},  {\em Astrophys. J} {\bf 155} (Jan., 1969)
  89.

\bibitem{1976ApJ...208L...1W}
S.~{Weinberg}, {\it {Apparent luminosities in a locally inhomogeneous
  universe}},  {\em Astrophys. J} {\bf 208} (Aug., 1976) L1--L3.

\bibitem{Sugiura:1999fm}
N.~Sugiura, K.-i. Nakao, D.~Ida, N.~Sakai, and H.~Ishihara, {\it {How do
  nonlinear voids affect light propagation?}},  {\em Prog. Theor. Phys.} {\bf
  103} (2000) 73--89, [\href{http://xxx.lanl.gov/abs/astro-ph/9912414}{{\tt
  astro-ph/9912414}}].

\bibitem{Kozaki:2002ka}
H.~Kozaki and K.-i. Nakao, {\it {Volume Expansion of Swiss-Cheese Universe}},
  {\em Phys. Rev.} {\bf D66} (2002) 104008,
  [\href{http://xxx.lanl.gov/abs/gr-qc/0208091}{{\tt gr-qc/0208091}}].

\bibitem{Brouzakis:2006dj}
N.~Brouzakis, N.~Tetradis, and E.~Tzavara, {\it {The Effect of Large-Scale
  Inhomogeneities on the Luminosity Distance}},  {\em JCAP} {\bf 0702} (2007)
  013, [\href{http://xxx.lanl.gov/abs/astro-ph/0612179}{{\tt
  astro-ph/0612179}}].

\bibitem{Marra:2007pm}
V.~Marra, E.~W. Kolb, S.~Matarrese, and A.~Riotto, {\it {On cosmological
  observables in a swiss-cheese universe}},  {\em Phys. Rev.} {\bf D76} (2007)
  123004, [\href{http://xxx.lanl.gov/abs/0708.3622}{{\tt arXiv:0708.3622}}].

\bibitem{Marra:2007gc}
V.~Marra, E.~W. Kolb, and S.~Matarrese, {\it {Light-cone averages in a
  swiss-cheese universe}},  {\em Phys. Rev.} {\bf D77} (2008) 023003,
  [\href{http://xxx.lanl.gov/abs/0710.5505}{{\tt arXiv:0710.5505}}].

\bibitem{Bolejko:2008xh}
K.~Bolejko, {\it {The Szekeres Swiss Cheese model and the CMB observations}},
  \href{http://xxx.lanl.gov/abs/0804.1846}{{\tt arXiv:0804.1846}}.

\bibitem{Vanderveld:2008vi}
R.~A. Vanderveld, E.~E. Flanagan, and I.~Wasserman, {\it {Luminosity distance
  in 'Swiss cheese' cosmology with randomized voids: I. Single void size}},
  {\em Phys. Rev.} {\bf D78} (2008) 083511,
  [\href{http://xxx.lanl.gov/abs/0808.1080}{{\tt arXiv:0808.1080}}].

\bibitem{Brouzakis:2007zi}
N.~Brouzakis, N.~Tetradis, and E.~Tzavara, {\it {Light Propagation and
  Large-Scale Inhomogeneities}},  {\em JCAP} {\bf 0804} (2008) 008,
  [\href{http://xxx.lanl.gov/abs/astro-ph/0703586}{{\tt astro-ph/0703586}}].

\bibitem{Brouzakis:2008uw}
N.~Brouzakis and N.~Tetradis, {\it {Analytical Estimate of the Effect of
  Spherical Inhomogeneities on Luminosity Distance and Redshift}},  {\em Phys.
  Lett.} {\bf B665} (2008) 344--348,
  [\href{http://xxx.lanl.gov/abs/0802.0859}{{\tt arXiv:0802.0859}}].

\bibitem{Gurzadyan:2008yx}
V.~G. Gurzadyan and A.~A. Kocharyan, {\it {Porosity criterion for hyperbolic
  voids and the cosmic microwave background}},  {\em Astron. Astrophys.} {\bf
  493} (2009) L61, [\href{http://xxx.lanl.gov/abs/0807.1239}{{\tt
  arXiv:0807.1239}}].

\bibitem{Gurzadyan:2008hj}
V.~G. Gurzadyan and A.~A. Kocharyan, {\it {Kolmogorov stochasticity parameter
  measuring the randomness in Cosmic Microwave Background}},
  \href{http://xxx.lanl.gov/abs/0810.3289}{{\tt arXiv:0810.3289}}.

\bibitem{Ghassemi:2009ug}
S.~Ghassemi, S.~K. Moghaddam, and R.~Mansouri, {\it {Lensing effects in
  inhomogeneous cosmological models}},
  \href{http://xxx.lanl.gov/abs/0901.0340}{{\tt arXiv:0901.0340}}.

\bibitem{Clifton:2009nv}
T.~Clifton and J.~Zuntz, {\it {Hubble Diagram Dispersion From Large-Scale
  Structure}},  \href{http://xxx.lanl.gov/abs/0902.0726}{{\tt
  arXiv:0902.0726}}.

\bibitem{Clifton:2009kx}
T.~Clifton, P.~G. Ferreira, and J.~Zuntz, {\it {What the small angle CMB really
  tells us about the curvature of the Universe}},
  \href{http://xxx.lanl.gov/abs/0902.1313}{{\tt arXiv:0902.1313}}.

\bibitem{Rees:1968zz}
M.~J. Rees and D.~W. Sciama, {\it {Large scale Density Inhomogeneities in the
  Universe}},  {\em Nature} {\bf 217} (1968) 511--516.

\bibitem{Granett:2008xb}
B.~R. Granett, M.~C. Neyrinck, and I.~Szapudi, {\it {Dark Energy Detected with
  Supervoids and Superclusters}},
  \href{http://xxx.lanl.gov/abs/0805.2974}{{\tt arXiv:0805.2974}}.

\bibitem{Masina:2008zv}
I.~Masina and A.~Notari, {\it {The Cold Spot as a Large Void: Rees-Sciama
  effect on CMB Power Spectrum and Bispectrum}},
  \href{http://xxx.lanl.gov/abs/0808.1811}{{\tt arXiv:0808.1811}}.

\bibitem{BMandelbrot:1983}
{B. B. Mandelbrot}, {\em The Fractal Geometry of Nature}.
\newblock {Macmillan}, {1983}.

\bibitem{Sachs:1961zz}
R.~K. Sachs, {\it {Gravitational waves in general relativity. 6. The outgoing
  radiation condition}},  {\em Proc. Roy. Soc. Lond.} {\bf A264} (1961)
  309--338.

\bibitem{Gorski:2004by}
K.~M. Gorski {\em et.~al.}, {\it {HEALPix -- a Framework for High Resolution
  Discretization, and Fast Analysis of Data Distributed on the Sphere}},  {\em
  Astrophys. J.} {\bf 622} (2005) 759--771,
  [\href{http://xxx.lanl.gov/abs/astro-ph/0409513}{{\tt astro-ph/0409513}}].

\bibitem{Einstein:1945id}
A.~Einstein and E.~G. Straus, {\it {The influence of the expansion of space on
  the gravitation fields surrounding the individual stars}},  {\em Rev. Mod.
  Phys.} {\bf 17} (1945) 120--124.


\end{thebibliography}\endgroup
\end{document}